\documentclass{aa}  

\usepackage{graphicx}
\usepackage{txfonts}
\usepackage{makecell}
\usepackage[colorlinks=true,urlcolor=blue,linkcolor=blue,citecolor=blue]{hyperref}
\begin{document} 

\newcommand{\dearwatson}{\textsc{dearwatson}\,}
\newcommand{\watson}{\textsc{watson}\,}
\newcommand{\kepler}{\textsc{Kepler}\,}
\newcommand{\tess}{\textsc{TESS}\,}
\newcommand{\ete}{\textsc{ETE-6}\,}
\newcommand{\dr}{\textsc{DR25}\,}
\newcommand{\lightkurve}{\textsc{Lightkurve}\,}
\newcommand{\wotan}{\textsc{Wotan}\,}
\newcommand{\exoml}{\textsc{exoml}\,}
\newcommand{\astropy}{\textsc{Astropy}\,}

   \title{WATSON-Net: Vetting, Validation, and Analysis of Transits from Space Observations with Neural Networks}

   %\subtitle{I. Overviewing the $\kappa$-mechanism}

   \author{Martín Dévora-Pajares
          \inst{1,2}
          \and
          F. J. Pozuelos\inst{3,6}
          \and
          J.C. Suárez\inst{1}\and
          M. González-Penedo\inst{4,6}\and
          C. Dafonte\inst{5,6}
          }

   \institute{Dpto. Física Teórica y del Cosmos. Universidad de Granada. 18071. Granada, Spain
         \and
             AI Engineering, Avature, Spain
         \and
             Instituto de Astrof\'isica de Andaluc\'ia (IAA-CSIC), Glorieta de la Astronom\'ia s/n, 18008
         \and
             CITIC-Research Center of Information and Communication Technologies, University of A Coruña, A Coruña, Spain
         \and
             CIGUS CITIC - Department of Computer Science and Information Technologies, University of A Coruña, A Coruña, Spain
         \and
             Artificial Intelligence for Research in Exoplanets and Stars (AIRExS), UDC, Associated Unit to CSIC through the IAA, 15008
             }

   \date{Received September 3, 2025; accepted X}

% \abstract{}{}{}{}{} 
% 5 {} token are mandatory
 
  \abstract
   {As the number of detected transiting exoplanet candidates continues to grow, the need for robust and scalable automated tools to prioritize or validate them has become increasingly critical. Among the most promising solutions, deep learning models offer the ability to interpret complex diagnostic metrics traditionally used in the vetting process.}
   {In this work, we present WATSON-Net, a new open-source neural network classifier and data preparation package designed to compete with current state-of-the-art tools for vetting and validation of transiting exoplanet signals from space-based missions.}
   {Trained on Kepler Q1–Q17 DR25 data using 10-fold cross-validation, WATSON-Net produces ten independent models, each evaluated on dedicated validation and test sets. The ten models are calibrated and prepared to be extensible for TESS data by standardizing the input pipeline, allowing for performance assessment across different space missions.}
   {For Kepler targets, WATSON-Net achieves a recall-at-precision of 0.99 (R@P0.99) of 0.903, ranking second, with only the ExoMiner network 
performing better (R@P0.99 = 0.936). For TESS signals, WATSON-Net emerges as the best-performing non-fine-tuned machine learning classifier, achieving a precision of 0.93 and a recall of 0.76 on a test set comprising confirmed planets and false positives. Both the model and its data preparation tools are publicly available in the \dearwatson Python package, fully open-source and integrated into the vetting engine of the SHERLOCK pipeline.}
   {}

   \authorrunning{Dévora-Pajares et al.}
   \titlerunning{WATSON-Net}
   \maketitle
%
%-------------------------------------------------------------------

\section{Introduction}
The vetting and analysis of transiting exoplanet candidate signals are essential components of any space-based mission dedicated to exoplanet discovery. As the volume of detected signals continues to grow, robust and automated vetting strategies have become increasingly critical for prioritizing follow-up efforts and ensuring the reliability of derived planet catalogs. The first missions, such as COROT \citep{corot2006}, released only a few dozen candidates and were primarily conceived as proof-of-concept efforts demonstrating what could be achieved with dedicated projects. Thus, vetting could be handled in manual, step-by-step pipelines. The subsequent mission, Kepler \citep{kepler2010}, was designed to enable the first large-scale statistical studies of exoplanet populations and included numerous automated processes across its multiple data pipelines. These steps would later be centralized by the Science Processing Operations Center \citep[SPOC,][]{spoc2016} prepared for the TESS \citep{tess2016} mission. After finding a signal using their Transiting Planet Search (TPS) pipeline \citep{tps2010}, every signal had to comply with several models, thresholds, and metrics computed with different techniques and algorithms, which were filtered one by one. Over time, the Kepler vetting pipeline underwent significant changes, and subsequent Data Releases (DRs) included only those signals that successfully passed the latest set of filters and metrics, excluding those that no longer met the updated criteria. To help categorize signals in a more comprehensible and simplified manner, several tools have been developed over the last decade, including Autovetter \citep{autovetter2015}, which utilizes a Random Forest classifier, and Robovetter \citep{robovetter2016}, a tool that transforms human decisions into machine resolution through simple decision trees. Other independent projects have emerged within the community, either aiming to prioritize potential exoplanet candidates \citep[see, e.g.,][]{diamante1}, or focusing on their vetting and validation, as demonstrated by the confirmed planets around Kepler-80 and Kepler-90 in \cite{kepler80}. In recent years, a completely new field has emerged in the data science and processing area with Neural Networks (NN) and their Deep Learning (DL) approaches, where several perceptron layers are connected between the data inputs and outputs, creating the so-called hidden layers with hidden units introduced in \cite{rumelhart1986}. Such approaches have taken over the state of the art of most scientific fields, such as Natural Language Processing \citep{devlin2019bert, touvron2023llama}, Time Series Analysis \citep{li2018recurrent, lopes2019prediction}, Image Processing/Computer Vision \citep{dosovitskiy2021vit}, and also a wide range of Astrophysics \citep{maier2024convolutional}. One of the first DL screening proposals was AstroNet \citep{astronet2019}, which was followed by several iterations of the same work using convolutional neural networks (CNNs), a DL architecture introduced in \cite{lecun1989}. This led to a final version with minor adjustments in \cite{astronet2019-1}. Soon, new approaches emerged, including additional information as the centroids shift data and the stellar parameters, as \cite{exonet} did with their Exonet NN. The latest and arguably most successful model in this line of development is ExoMiner \citep{exominer2022}, which proposed a multibranch NN architecture and demonstrated superior performance over all previous methods. ExoMiner not only proved to be the most performant model under the same test sets compared to the previously published tools but was also used to release more than 300 new, validated transiting exoplanets from the Kepler Data Release 25 catalog \citep[DR25,][]{dr25_2016}. While modern methods based on DL techniques have shown strong potential in validating planetary candidates from space missions, approaching these results cautiously is essential, especially when there is no independent confirmation from follow-up efforts. Additionally, ExoMiner was also enhanced with new training on TESS data, reaching state-of-the-art results for this mission \cite[ExoMiner++;][]{exominer2025}. 

In this context, here we present WATSON-Net, a DL–based tool designed to perform automated vetting of transiting exoplanet candidates. WATSON-Net adopts a multi-branch NN architecture to classify signals as either True Positives or False Positives, returning a probabilistic score that reflects the likelihood of a planetary origin. To facilitate its practical use in ongoing surveys, WATSON-Net has been integrated into the latest version of the SHERLOCK pipeline, a modular, open-source framework for detecting and analyzing planetary candidates in light curves from space-based missions \citep{devora2024, pozuelos2020}. While earlier versions of SHERLOCK relied on manual inspection of heuristic diagnostics through a preliminary module named \watson, the inclusion of WATSON-Net enables a fully automated and interpretable vetting process, significantly reducing the effort required for signal assessment.

We further benchmark WATSON-Net against leading models in the field, including ExoMiner, focusing on their performance with Kepler and TESS data. In particular, we analyze their ability to reproduce confirmed exoplanet populations and discuss the implications for the robustness and reliability of modern DL approaches in the context of space-based transit surveys.

\section{Methodology}
Training, validation, and test set selection and preparation are among the most critical stages of any Machine Learning (ML) model. The choice of data to be used in the entire training process is crucial for obtaining reliable predictions when applied in productive environments with inputs that were never seen during the training steps. Therefore, the training data must, whenever possible, be well-curated by experts from the community on which the model is focused. In addition, it is not only necessary to start with a robust and trustworthy labeled dataset, but also to prepare the data in a clean and numerically stable manner to facilitate the model's ability to distinguish between scenarios. We decided to rely exclusively on data available through public web services and to provide the data processing as part of the final release to be able to compute the same metrics as those provided for the official Threshold-Crossing Events (TCEs) list \citep{jenkins2017}. This approach ensures that all the input data for WATSON-Net is locally computed, avoiding dependency on the publication of official candidate lists with their own metrics, and allowing users to obtain predictions for any signal of their choice.

\subsection{Datasets selection}
To ensure high data quality, we used the latest data release (DR25) of TCEs and Kepler Objects of Interest (KOIs)\footnote{\url{https://exoplanetarchive.ipac.caltech.edu/cgi-bin/TblView/nph-tblView?app=ExoTbls\&config=tce}}. This catalog reflects several improvements and iterations on the SPOC pipeline, each enhancing its performance and yielding a robust, well-labeled dataset. In addition, we extended the testing of our model to evaluate its generalization to \tess data by incorporating a subset of known and confirmed planets, as well as false positives, from the official TESS Objects of Interest (TOIs) catalog. These \tess signals are broadly comparable to those in DR25, as they are also classified using SPOC algorithms and metrics. At the same time, the larger pixel scale of \tess (21 arcsec versus Kepler’s 4 arcsec) introduces a different observational context, particularly in terms of background source blending. As a result, some diagnostic metrics must be interpreted with greater care, while still providing valuable information for vetting transit-like signals in \tess light curves.

\subsection{Neural network design}
\label{sec:nn_design}
Many of the state-of-the-art NNs are based on using Transformers \citep{vaswani2017} nowadays. Transformers are, however, challenging to train because they require substantial amounts of data to generalize, that is, to prevent overfitting to the training data. Additionally, their training requires more powerful computing infrastructure. Therefore, we continued using CNNs, as they often report remarkable results for classification tasks involving time series, due to their ability to extract both local and global features from time-related data, as demonstrated in \cite{zhao_cnns}. 

The inputs to our network were carefully selected based on domain knowledge, past results in the literature, and empirical performance. Each of the following components aims to provide relevant information for distinguishing genuine planetary transits from false positives. Specifically, we use:

\begin{enumerate}
    \item A \textbf{global phase-folded light curve}, binned into 300 bins, centered at the signal's epoch. This input captures the overall transit shape across the full orbital phase. The choice of 300 bins balances temporal resolution and noise robustness, based on prior experience using other validation tools like TRICERATOPS \cite{giacalone2021}. Fewer bins (e.g., 100) reduced resolution of the transit shape, specially for short transits, and more bins increased the training cost at a questionable increment of the information provided by this curve. Centering at the epoch ensures that the transit occurs near the same position in all training examples, which helps the convolutional filters learn consistent features related to the transit signal.
    
    \item \textbf{Four phase-folded light curves}, each binned into 75 bins and focused on a window six times the transit duration. These include: 1) the odd transits, 2) the even transits, 3) the folded curve focused on the transit epoch, 4) the folded curve focused on the expected location of the secondary eclipse. This localized zoom-in allows the network to learn subtle differences in transit depth and shape (e.g., due to eclipsing binaries), as well as any significant secondary event. The window size of six transit durations (2.5 before and after, plus the transit itself) was chosen to capture both the full transit profile and potential nearby anomalies. A narrower window (e.g., 3 durations) sometimes excluded close-in variability, while a broader window added redundant out-of-transit noise.
    
    \item A \textbf{centroid shift time series}, binned into 75 points and centered on the transit, again using a window of six transit durations. This input helps the network detect any significant motion of the flux centroid, which could indicate a background eclipsing binary rather than a planet. The same windowing rationale as above applies here — we want to observe motion directly associated with the transit-like event, without diluting the signal with unrelated motion before or after.

    \item A \textbf{binned optical ghost time series} \citep{dvr1}, also focused on a six-transit-duration window. Ghost signals are a known source of false positives in space-based photometry. Providing this input enables the model to learn patterns associated with instrumental or optical artifacts.

    \item A set of \textbf{numerical transit parameters}, including:
    \begin{itemize}
        \item Period, duration, planetary radius,
        \item Number of transits with sufficient coverage,
        \item Number of high-quality transits (with at least 75\% of in-transit points available),
        \item Statistical vetting metrics such as the bootstrap false alarm probability \citep{jenkins2017}, secondary event albedo, secondary event temperature and odd-even factor.
    \end{itemize}
    These features provide a statistical and physical summary of the signal quality and consistency. While some of this information could be inferred from the light curves, including them explicitly improves learning and interpretability.

    \item The \textbf{source offset significance} \citep{dvr1}, which measures whether the transit occurs at the position of the target star or is offset, as could happen with blended background binaries. This is a high-level vetting metric used in expert pipelines, and gives the model direct access to spatial information derived from image differencing.

    \item Finally, we include the \textbf{stellar radius, mass, and effective temperature} of the host star, derived from stellar characterization catalogs. These values help the network understand whether the inferred planetary radius is physically plausible. Although these parameters have associated uncertainties, we use their central values only, since including the errors did not lead to improved performance in preliminary tests. A future improvement could incorporate posterior samples or uncertainty-aware inputs, but we found the central estimates sufficient for classification.
\end{enumerate}

The global flux input is fed into a convolutional branch with four groups of convolutional layers, as shown in Fig.~\ref{fig:global-flux-conv}. All the convolutional and dense layers are activated with a leaky ReLU \citep{leakyReLU} activation function to ensure the gradient does not become zero for small negative values. A Gaussian noise initial layer is used to create variability on the inputs and simulate oversampling with a simple and effective approach. The main and secondary events, odd and even, centroids, and optical ghost-focused flux inputs are fed into a generalized convolutional branch represented in Fig.~\ref{fig:watson-conv-branch}. As in the global flux convolutional branch, all the convolutional and dense layers also utilize a leaky ReLU activation function. The same generalized convolutional branch is used for the odd and even focused flux data. The transit numerical parameters are fed into the main and secondary events branches, and the source offset significance value is injected into the centroids branch. All these convolutional branches ultimately pass through a layer named Branch Dropout, whose purpose is to nullify the entire branch (set values to zero) if a random float value between 0 and 1 exceeds a given threshold. With this approach, we aim to train the NN on scenarios where any of the branches may not be present, thereby enhancing the potential for future explainability cases. 

The global flux input is fed into a convolutional branch with four groups of convolutional layers, as shown in Fig.~\ref{fig:global-flux-conv}. All convolutional and dense layers are activated with a leaky ReLU function. A Gaussian noise initial layer is included to introduce controlled variability into the inputs, effectively simulating oversampling in a simple yet effective manner. The main and secondary events, odd and even, centroids, and optical ghost-focused flux inputs are processed through a generalized convolutional branch (Fig.~\ref{fig:watson-conv-branch}), which follows the same activation strategy as the global flux branch. The same generalized convolutional branch configuration is used for both the odd and even focused flux data to ensure parameter sharing and consistent feature extraction across these comparable inputs. The transit numerical parameters are fed into the main and secondary event branches, while the source offset significance value is injected into the centroids branch.

This multi-branch architecture was chosen to allow each input type to be processed in a domain-specific manner before combining their learned features. Light curves at different time scales, centroid time series, and optical ghost indicators exhibit distinct temporal and noise characteristics, which benefit from dedicated convolutional filters tuned to each data type. We explored several alternative configurations during development, including deeper convolutional stacks, varying kernel sizes, and different numbers of filters per layer. Architectures with more convolutional layers tended to overfit the training set without yielding improvements in validation performance, while shallower networks showed reduced sensitivity to subtle transit-like features. The final configuration of four convolutional groups for the global flux branch and the standardized generalized branch for the focused and auxiliary time series was selected as the optimal balance between model complexity, training stability, and generalization ability, as determined by empirical evaluation on validation data.

All convolutional branches ultimately feed into a layer termed Branch Dropout, which nullifies an entire branch (setting its values to zero) if a uniform random float between 0 and 1 exceeds a given threshold. This mechanism forces the network to learn to classify signals even when one or more diagnostic inputs are absent, improving robustness and enabling future explainability analyses in cases where incomplete data may be available.

\begin{figure*}
    \centering
    \includegraphics[width=0.95\textwidth]{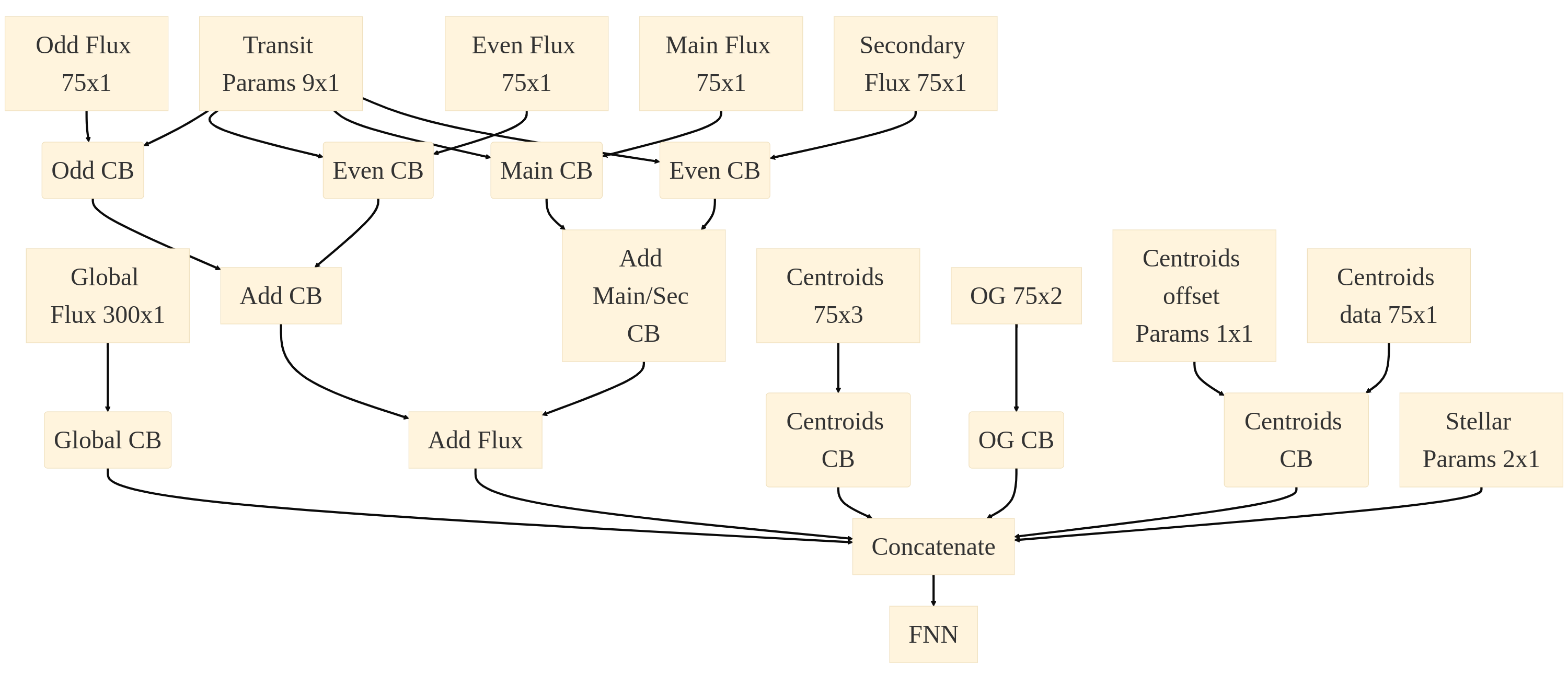}
    \caption{Neural network global architecture.}
    \label{fig:watson-net}
\end{figure*}

The perceptron weights of the odd, even, main, and secondary event flux branches are all added together. Later, all the resulting branches are merged through a concatenation layer, which is then passed through a Feed-Forward Network (FFN), as shown in Fig.~\ref{fig:watson-ffn}. The entire network is shown in Fig.~\ref{fig:watson-net}.

% %%{ init: {
%     'flowchart': { 'curve': 'stepBefore' },
%     'theme': 'base',
%     'themeVariables':
%         { 'fontSize': '30px', 'fontFamily': 'Inter'}
% } }%%
%  flowchart TD
%      A1[Transit Params 9x1] --> D1(Odd CB)
%      A1[Transit Params 9x1] --> E1(Even CB)
%      A1[Transit Params 9x1] --> F1(Har. Odd CB)
%      A1[Transit Params 9x1] --> G1(Subhar. Odd CB)
%      D[Odd Flux 75x1] --> D1(Odd CB)
%      E[Even Flux 75x1] --> E1(Even CB)
%      F[Main Flux 75x1] --> F1(Main CB)
%      G[Secondary Flux 75x1] --> G1(Secondary CB)
%      G2[Global Flux 300x1] --> G3(Global CB)
%      G3(Global CB) --> L[Concatenate]
%      D1(Odd CB) --> H[Add CB]
%      E1(Even CB) --> H[Add CB]
%      F1(Main CB) --> I[Add Main/sec CB]
%      G1(Even CB) --> I[Add Main/sec CB]
%      J[Centroids 75x3] --> J1(Centroids CB)
%      K[OG 75x2] --> K1(OG CB)
%      H[Add CB] --> II[Add Flux]
%      I[Add Main/Sec CB] --> II[Add Flux]
%      II[Add Flux] --> L[Concatenate]
%      J1(Centroids CB) --> L[Concatenate]
%      A2[Centroids offset Params 1x1] --> M[Centroids CB]
%      A3[Centroids data 75x1] --> M[Centroids CB]
%      A4[Stellar Params 2x1] --> L[Concatenate]
%      M(Centroids CB) --> L[Concatenate]
%      K1(OG CB) --> L[Concatenate]
%      L[Concatenate] --> L1[FNN]

\subsection{Data curation}
Pre-processing the selected dataset to be used as input by any ML algorithm is a fundamental stage that is critical to developing a working and successful model. This is especially true for DL models, where the computed gradient can vanish or explode easily for very low, huge, or problematic input values. Our neural network model was designed to handle both single numerical inputs (e.g., stellar radius, effective temperature) and time series data (e.g., flux, centroid offsets).
The initial step involved curating and systematically classifying the dataset to ensure the removal of duplicates and assigning consistent labels to each entry. Each signal was categorized as either a Transiting Planet (TP) or a Non-Transiting Planet (NTP), based on a set of rigorous selection criteria. The specific steps undertaken in this classification process are detailed below:

\begin{itemize}
    \item Cross-checked Kepler Q1-Q17 DR25 KOI and TCE data with the Planetary Systems Composite data and the Kepler Certified False Positive data.
    \item Removed any entry classified as rogue transits from the dataset.
    \item KOIs found from old data releases (e.g., Q1-Q17 DR24) were excluded from the dataset.
    \item In cases where multiple TCEs or KOIs were detected for the same target and exhibited period values within 5\% of each other, all corresponding signals were discarded. Such proximity in period and epoch often leads to mutual contamination during phase-folding, as each signal introduces artifacts into the light curve of the other, thereby degrading the quality of the training data.
    \item Any TCE that did not match any KOI was set as NTP.
    \item Pre-computed all the inputs for each target to be used during training.\citep{wotan}, assuming an optimal window size of four times the signal transit duration.
    \item Any entry disposed as certified FP or certified False Alarm (FA) in the Kepler Certified False Positive catalog was labeled as NTP.
    \item Any remaining KOI whose disposition was not listed as confirmed was labeled as a candidate and excluded from both the training and validation sets in order to avoid introducing ambiguous or uncertain classifications during model training.
    \item Signals classified as confirmed in previous machine-learning-based validation efforts \citep{2021_gpc, exominer2022, exominer2023} were relabeled as candidates and excluded from the training and validation sets. These signals were intentionally set aside for testing purposes, as they had already been used to validate other neural network models, and including them could bias the performance evaluation of our own approach.
    \item Any remaining signal with a confirmed disposition was labeled as TP.
\end{itemize}

\begin{figure}
    \centering
    \includegraphics[width=\columnwidth]{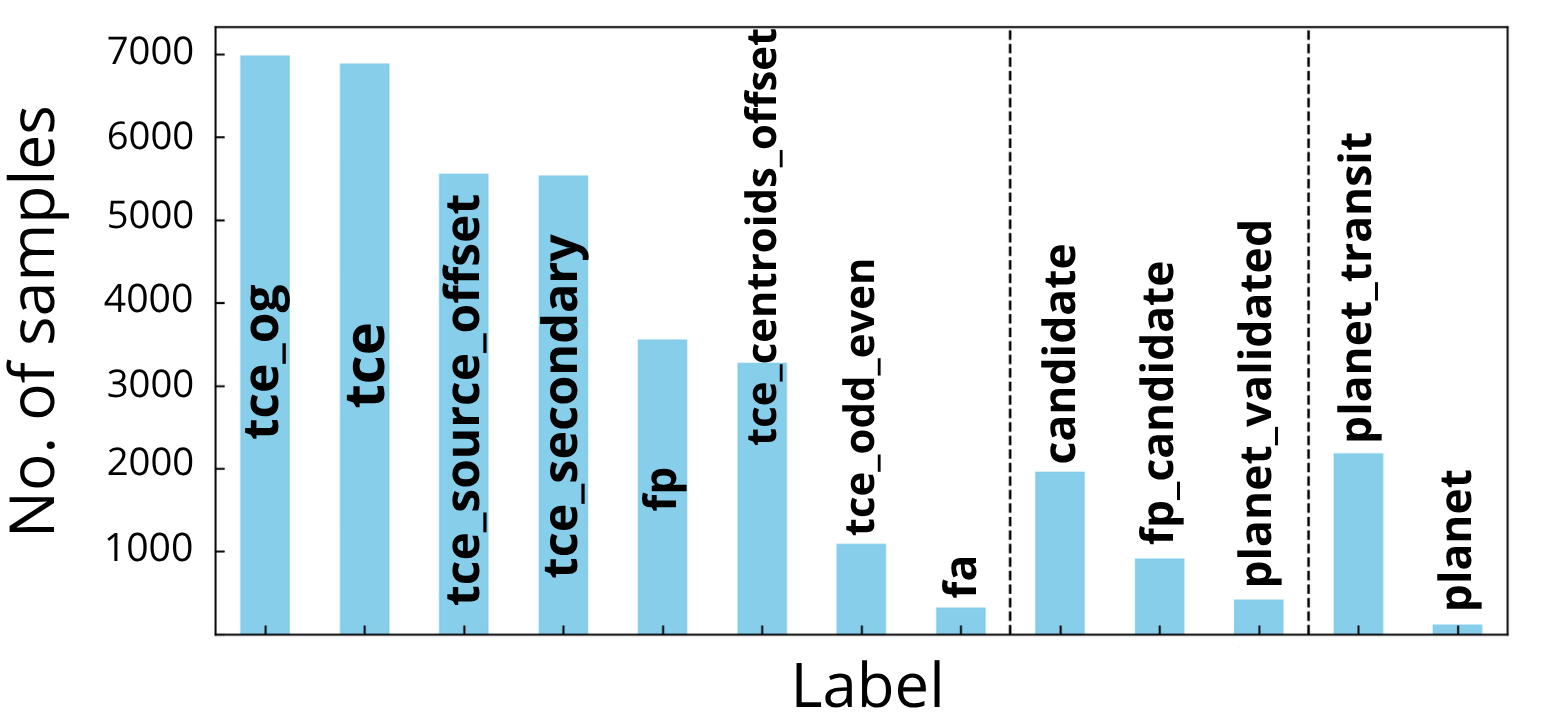}
    \caption{Number of signals matching each of the label criteria. Each NTP signal can potentially match different labels. Dashed vertical lines split three different regions: the left contains the NTP-labeled samples, the middle contains the signals considered as candidates, and the right contains signals labeled as TP (see text for details).}
    \label{fig:dataset-tags}
\end{figure}

Even when all the negative, candidate, and positive samples were marked with either NTP, candidate, or TP tags, each was cataloged with such a disposition for multiple reasons. Negative samples could be rejected because of optical ghost effects (tce\_og), significant transit source offset (tce\_source\_offset), official FP disposition (fp), significant centroids offset (tce\_centroids\_offset), significant odd/even effect (tce\_odd\_even), or official FA disposition (fa). KOIs with official FP disposition but not present in the Kepler Certified False Positive catalog were labeled with fp\_candidate. TCEs marked as candidates were tagged similarly, and validated planets excluded from the training set were classified as planet\_validated. Under the TP classification, we distinguished between transiting planets (planet\_transit) and planets only confirmed using radial velocities (planet). The distribution of these fine-grained classifications is illustrated in Fig.~\ref{fig:dataset-tags}. As a final result, a dataset of 23966 targets was obtained for the DR25 catalog. 

Subsequently, we prepared the input data for the NN. Using the \lightkurve package \citep{lightkurve}, we retrieved the Q1–Q17 short- and long-cadence light curves, along with the Target Pixel Files (TPFs) for each Kepler target. Stellar parameters were obtained from the NASA Mikulski Archive for Space Telescopes (MAST). Once all raw data were collected, each input type underwent an independent preprocessing procedure, as detailed below:
\begin{itemize}
    \item The transit parameters were comprised of the normalized values:
    the normalized transit duration, whose original value we divided by 15 (as it was close above the maximum transit duration in our dataset); the normalized planet radius, whose original value we divided by 300 (close above the maximum transit radius in our dataset); the good transits count, which is the count of all the single transits with data covering more than 75\% of the transit duration; the good transit count normalization (setting a limit of 20 transits as a sufficient number of transits to consider a signal reliable enough); and the good transit ratio, which is the good transit count divided by the total number of single transits with at least one data point within the transit duration.
    \item As aforementioned, there were four statistical metrics. The bootstrap FAP is computed obtaining the Signal Detection Efficiency (SDE) of the signal as defined by \cite{tls}. The complete light curve is then randomly permuted, and a new search is run again, storing the highest SDE. The permutation and search are repeated 100 times, and the percentage of SDEs obtained from random permutations above the original signal SDE is used as bootstrap FAP. The secondary event albedo and temperature statistics are computed similarly to \cite{jenkins2017}. The odd/even factor is obtained by normalizing the absolute value of the even and odd event depths.
    \item Only two stellar parameters are used: $T_{eff}$ and $R_s$.
    \item We begin with the Pre-Search Data Conditioning Simple Aperture Photometry (PDCSAP) fluxes, which have undergone preliminary corrections for instrumental systematics and crowding effects \citep{stumpe2012,smith2012}. Any remaining unnormalized fluxes are normalized to ensure consistency across the dataset. To mitigate long-term variability and systematic trends that may obscure planetary transit signals, we applied a bi-weight filter with a window size set to four times the transit duration, using the implementation provided by the \texttt{WOTAN} package \citep{hippke2019}. This detrending step enables us to remove intrinsic stellar activity, instrumental noise, and drift while preserving the transit signal. Following this correction, we extract the relevant transit features by constructing phase-folded light curves for global, primary, and secondary transit events, as well as for odd and even transit separations to aid in identifying eclipsing binaries. All flux values are then normalized by dividing by 2, ensuring that the data is consistently scaled within the range [0,1] and centered at 0.5, which stabilizes the training process of the neural network. This normalization step is crucial for mitigating numerical instabilities, such as gradient vanishing or explosion, which could otherwise degrade model convergence and performance.
    \item We examined the TPFs to calculate the shift in centroids for both right ascension and declination. To do so, the shift needs to be computed first for each TPF in pixel units with the following expressions:

    \begin{gather*}       
        C_i=\frac{c_i - median(c_{i,oot})}{std(c_i)} \\
        M_i=\frac{m_i - median(m_{i,oot})}{std(m_i)} \\
        S_i=M_i - C_i
    \end{gather*}

    Where $c_i$ is the centroid position on each cadence for a given row or column, $m_i$ is the measured motion of the focal plane respectively to the target position, $c_{i,oot}$ are the centroid positions outside the transit signal, $m_{i,oot}$ are the motion measured outside the transit signal, $C_i$ is the final computed centroid shift for each point on a given row or column, $M_i$ is the final computed motion for each point on a given row or column. $S_i$ is the final shift computed by subtracting the centroid deviation position from the computed motion for rows and columns separately. Finally, using the TPF data and the \astropy package \citep{astropy1, astropy2}, we convert the row and column shifts to right ascension and declination shifts for each TPF.
    \item In the case of the optical ghost, we also run the computations for each TPF separately. A halo aperture is calculated around the official pipeline mask by applying a dilation and a simple AND logical operation. Then, the core aperture mean flux is computed with the average value of all the pixels within the pipeline aperture. The same computation is applied to the halo flux for points within the halo aperture pixels. Both fluxes are then detrended with a bi-weight filter and a window size of four times the signal transit duration.
    \item The source offset and its error are computed using the average from the values obtained with the proposed methods in Sec.~2.3.4 and 2.3.5 from \cite{devora2024}. The offset significance is then calculated and normalized by dividing the computed source offset error by twice its value.
\end{itemize}

In any case, where the normalized parameters exceeded 1 or 0, we truncated them and set them to $1 - 10^{-7}$ or $10^{-7}$, respectively. With the exposed processing, the data was curated, and data preparation was completed before the training stage began.

\subsection{Neural network training setup}

\subsubsection{Hyperparameters exploration}
Hyperparameter optimization is one of the most challenging aspects of NN training, typically addressed through automated tools that explore predefined grids and return the configuration yielding the best metric. However, such searches are only indicative, since promising early results do not guarantee long-term improvement. The number of hyperparameters depends strongly on the model architecture, with layer number and size being common targets. Importantly, sequential optimization (varying one parameter at a time) is generally insufficient due to nonlinear interdependencies, where small changes can significantly affect model behavior \citep{bergstra2012}. Thus, effective tuning requires holistic strategies. In our case, after exploratory experiments, we defined a specific set of hyperparameters to be tuned, as detailed below.

\begin{itemize}
    \item Batch size: Controls samples per update, affecting complexity, generalization, and imbalance handling. 
    \item Epochs: Full passes over the dataset; optimized via early stopping to avoid overfitting. 
    \item Optimizer and learning rate: Parameters updated with Adam \citep{2014_adam} ($\beta_1=0.9$, $\beta_2=0.98$, $\epsilon=1\times 10^{-9}$); learning rate scheduled with warm-up and decay. 
    \item Learning rate progression: Scaling factor applied at selected layers to improve convergence. 
    \item Gradient clipping: Prevents vanishing/exploding gradients \citep{bengio1994} by bounding norm or value. 
    \item Class sampling: Adjusts sample counts per class to balance binary dataset with diverse negatives. 
    \item Class loss weights: Loss modifiers to mitigate class imbalance. 
    \item Early stopping: Stops training once validation loss degrades. 
    \item Loss function: Guides optimization; tested Binary Cross-Entropy \cite{bce_loss}, Focal Loss \cite{focal_loss}, and Dice Loss \cite{dice_loss}. 
    \item Dropout (standard/spatial): Prevents overfitting by random connection removal \citep{2014_dropout,2014_spatialdropout}. 
    \item Branch dropout: Randomly drops entire branches during training. 
    \item White noise: Gaussian noise added at inputs to improve generalization. 
    \item L1/L2 regularization: Penalizes weights (L1 sparsity, L2 small values), applied separately to conv/dense layers. 
    \item SWA: Stochastic Weight Averaging \citep{2018_swa1,2018_swa2} reduces overfitting by averaging weights after fixed epochs. 
\end{itemize}

\subsubsection{Dataset slicing}
For the training phase, we adopted a standard 90\% -10\% split between the training and validation sets. The validation set consists of a subset of the original data exclusively used for performance evaluation at the end of each training epoch. Importantly, no validation data is used in the gradient computation, ensuring that it does not influence the weight updates in any NN layer.
Following this partitioning, the training set comprises 21,569 signals, while the validation set contains 2,397 signals.

\subsubsection{Cross Validation}
The stochastic nature of NN training risks local minima and overfitting, particularly with small datasets. To mitigate this, cross-validation ensures robustness and generalization. Following \cite{exominer2022}, we applied 10-fold cross-validation \citep{kohavi1995}, training on 90\% of the data while validating on disjoint 10\% subsets. This strategy entails training the model 10 times, where, in each iteration, a different, non-overlapping 10\% subset of the data is designated as the validation set, while the remaining 90\% is used for training. Hence, we leverage the full dataset, revealing performance variations under dataset imbalances, and allowing either selection of the best model or ensemble averaging across folds. 

\subsection{Metrics}
To monitor the model’s performance throughout training, we computed several evaluation metrics at each epoch. Among these, it is essential to identify and prioritize a reference metric that guides model selection, facilitates the creation of training checkpoints, and enables the application of early stopping criteria when signs of overfitting or numerical instabilities are detected. Specifically, we tracked:
\begin{itemize}
    \item Precision ($\frac{TP}{TP + FP}$): Measures the proportion of true positives among all signals classified as positive by the model. It reflects how reliable the model's positive predictions are.
    \item Recall ($\frac{TP}{TP + FN}$): Measures the proportion of true positives that the model successfully identifies among all real positive cases. It indicates the model’s ability to detect actual planets.
    \item Accuracy ($\frac{TP + TN}{TP + FP + TN + FN}$): Measures the overall proportion of correct predictions, including both true positives and true negatives. While commonly reported, accuracy can be misleading in highly imbalanced datasets like ours, where the number of real planets is much smaller than the number of false positives.
    \item Recall at a given precision (0.99, 1.00): Recall and precision are inherently linked and often exhibit a trade-off as the decision threshold varies; raising one typically leads to a decrease in the other. Since our primary objective is to maintain high confidence in the model’s positive predictions, we fix the precision at a predefined value (e.g., 0.99) and compute the recall at the threshold where this precision is reached. This metric quantifies the number of real planets the model can recover while ensuring that at least 99\% of its positive predictions are correct.
    \item Precision at K (P@K): This metric assesses the proportion of the model top \( K \) predictions that correspond to true positives. Given a test set, the predictions are ranked by their confidence scores, and the fraction of the top \( K \) entries that are correct is computed. Precision at \( K \) is particularly useful for assessing the model’s ability to prioritize the most promising candidates and is used exclusively during the evaluation phase, not during training.
    \item Negative Predictive Value (NPV, $\frac{TN}{TN + FN}$): Measures the proportion of true negatives among all cases the model predicts as negative. Hence, it reflects how often the model is correct when it predicts that a signal is not a planet. This metric complements precision by evaluating the model’s reliability in ruling out candidates.
    \item Specificity ($\frac{TN}{TN + FP}$): Measures the proportion of true negatives correctly identified among all real negative cases. Together with the NPV, this metric helps assess how well the model performs when predicting that a signal is not a planet.
    \item Specificity at a given NPV (0.999, 1.0): Analogous to Recall at a given Precision, this metric evaluates the specificity the model can achieve when maintaining a fixed high value of negative predictive value (NPV), such as 0.999. It allows us to determine how confidently the model can rule out candidates while still correctly identifying a large fraction of the true negatives. This is especially useful when analyzing predictions with very low probabilities of being planets.
\end{itemize}

All these metrics, which are illustrated in Table~\ref{tab:metrics}, are the primary reported information from the training and testing sets, which will be used to assess our model performance and will be utilized throughout the rest of our work.

\begin{table*}
\centering
\caption{Summary of evaluation metrics used in WATSON-Net. Metrics are grouped by type, with definitions, purposes, and intuitive examples.}
\begin{tabular}{p{1.8cm}p{4cm}p{4.2cm}p{6.0cm}}
\hline
\textbf{Metric} & \textbf{Definition} & \textbf{Purpose} & \textbf{Example / Intuition} \\
\hline
\multicolumn{4}{l}{\textbf{1. Basic Classification Metrics}} \\
Precision & \( \frac{\mathrm{TP}}{\mathrm{TP} + \mathrm{FP}} \) & Correctness of predicted positives. & Out of 100 predicted planets, 95 are true → Precision = 0.95. \\
Recall & \( \frac{\mathrm{TP}}{\mathrm{TP} + \mathrm{FN}} \) & Retrieval ratio of real positives. & Model detects 90 of 100 real planets → Recall = 0.90. \\
NPV & \( \frac{\mathrm{TN}}{\mathrm{TN} + \mathrm{FN}} \) & Correctness of predicted negatives. & 99 of 100 predicted non-planets are correct → NPV = 0.99. \\
Specificity & \( \frac{\mathrm{TN}}{\mathrm{TN} + \mathrm{FP}} \) & Retrieval ratio of real negatives. & 900 of 950 false positives are flagged → Specificity $\sim$ 0.947. \\
Accuracy & \( \frac{\mathrm{TP} + \mathrm{TN}}{\mathrm{Total}} \) & Correctness of all predictions. & 950 out of 1000 predictions are correct → Accuracy = 0.95. \\
\hline
\multicolumn{4}{l}{\textbf{2. Threshold-Based Metrics}} \\
R@P0.99 & Recall at precision = 0.99 & Recal at high precision & Recovery of real planets when 99\% of true predictions are correct. \\
T@P0.99 & Threshold for precision = 0.99 & Threshold above which positives are safe. & Above threshold~=~0.646, 99\% of the predicted signals are real positives. \\
R@P1.0 & Recall at precision = 1.0 & Recall at perfect precision. & Recovery of real planets with perfect precision. \\
T@P1.0 & Threshold for 1.0\% precision & Minimum threshold for zero false positives. & Above threshold~=~0.961, all predicted signals are real positives. \\
S@NPV0.999 & Specificity at NPV = 0.999 & Sensitivity at high NPV. & 99.6\% of the negative samples are properly labeled when 99.9\% of negative predictions are correct. \\
T@NPV0.999 & Threshold for NPV = 0.999 & Threshold below which negatives are safe. & Predictions below threshold~=~0.056 are 99.9\% true negatives. \\
S@NPV1.0 & Specificity at NPV = 1.0 & Sensitivity at perfect NPV. & 85.4\% of the negative samples are properly labeled when all the negative predictions are correct.\\
T@NPV1.0 & Threshold for NPV = 0.999 & Maximum threshold for zero false negatives. & Predictions below threshold~=~0.056 are real negatives. \\
\hline
\multicolumn{4}{l}{\textbf{3. Ranking-Based Metrics}} \\
P@K500 & Precision over top 500 predictions & Quality of most confident predictions. & 499 of top 500 are correct → P@K500 = 0.998. \\
\hline
\multicolumn{4}{l}{\textbf{4. Calibration Metrics}} \\
ECE & Expected calibration error & Assess model probabilities deviation from expectancy & Weighted average on fraction of positives deviation from score real probability \\
MCE & Max calibration error & Worst deviation of among prediction confidences and real probability. & The fraction of positives for 0.97 scores is 0.66 against expected 0.97 (MCE~=~0.21). \\
\hline
\end{tabular}
\label{tab:metrics}
\end{table*}

\subsection{Model calibration}
\label{sec:model_calibration}
The evaluation of a machine classifier performance cannot rely solely on a selected metric dependent on a threshold, as its selection significantly influences the correlation between the predicted probabilities of the classifier and the expected probabilities of the observed outcomes. Whenever a classifier is designed to be used as a decision-making tool (e.g., to validate new exoplanets), it is essential to assess its calibration using reliability diagrams. These graphical expressions assess the agreement of predictions and the value expectancy, allowing the classifier's error rate to be established. This calibration procedure is extensively explored in the current landscape of the field \citep[see, e.g., ][]{2021_gpc,exominer2022}. To construct such reliability diagrams, the range of predicted probabilities is partitioned into discrete bins. The mean predicted probability and the corresponding empirical probability are computed for each bin. Finally, these probabilities are plotted to inspect the deviation of the model's predictions from their expected likelihood, a concept introduced by \cite{reliabilitydiagram1983} in the context of weather forecasting. 
In this study, we explore two distinct methods. 

On the one hand, we employed the isotonic regression \citep{isotonic1972, isotonic2002} to enforce a monotonic relationship between the predicted probabilities and their calibrated counterparts. This non-parametric approach does not assume any predefined functional form; instead, it performs a piecewise mapping that preserves the ranking of the original predictions. Consequently, the calibrated probabilities exhibit a step-wise structure, with local adjustments based on the empirical distribution of the data.

On the other hand, we employed Platt scaling \citep{platt2000}, by fitting a logistic regression model to transform the raw model's outputs into the final calibrated probability estimates. Unlike isotonic regression, Platt scaling imposes a sigmoidal shape on the probability transformation, which can be particularly beneficial for models whose predictions exhibit a systematic bias toward extreme values.

There are other numerical tools available to estimate the errors in model calibration. In particular, we utilize the Expected Calibration Error (ECE) and Maximum Calibration Error (MCE), as introduced in \cite{ecemce2015} and further elaborated upon, especially in the context of NN use cases, in \cite{ecemce2017}. These metrics enable direct comparisons of calibration performance across different models evaluated on the same dataset. The ECE is computed based on the weighted average of the calibration errors for each calibration bin, whereas the MCE reports the maximum deviation from the predicted likelihood for any of the bins, capturing the worst-case calibration error.
   
\section{Model construction and results}\label{sec:model}
\subsection{Training}
After manually exploring a wide range of hyperparameters, we selected the final values listed in Table~\ref{tab:hyperparameters}. In addition, to simplifying the network and avoiding the dependency on numerical values like the transit or stellar parameters, we used a hyperparameter to flag their usage or removal. We appreciated a significant drop of 50\% in the R@P0.99 obtained with the same network, including these parameters. Thus, their use proved to be significant, and we kept them in the final NN design. We launched cross-validation training for every validation cluster and retried all of them several times to obtain the best result for each. The training took approximately 5 hours per trial and slice, totaling approximately 2 days of training on a low- to medium-sized GPU with 8 GB of memory and 800 CUDA cores. The summary of the metrics used to select the final models is listed in Table~\ref{tab:cv-metrics-stats} (Table~\ref{tab:cv-metrics} provides fine-grained information). 

\begin{table*}
    \centering
    \caption{Metrics computed for each of the trained cross-validation folds and their mean, median and standard deviation.}
    \label{tab:cv-metrics-stats}
    \begin{tabular}{lcccccccr} 
        \hline  
        CV Fold No. & R@P0.99 & T@P0.99 & R@P1.0 & T@P1.0 & S@NPV999 & T@NPV999 & S@NPV1.0 & T@NPV1.0\\
         \hline
        \multicolumn{9}{c}{Uncalibrated model} \\
         \hline
          Mean & 0.903 & 0.736 & 0.623 & 0.912 & 0.876 & 0.072 & 0.459 & 0.015\\
          Median & 0.918 & 0.739 & 0.712 & 0.982 & 0.975 & 0.035 & 0.399 & 0.004\\
          $\sigma$ & 0.051 & 0.195 & 0.220 & 0.127 & 0.292 & 0.081 & 0.462 & 0.029\\
         \hline
        \multicolumn{9}{c}{Isotonic calibrated model} \\
        \hline
          Mean & 0.894 & 0.607 & 0.622 & 0.965 & 0.872 & 0.097 & 0.709 & 0.023\\
          Median & 0.903 & 0.646 & 0.715 & 0.961 & 0.966 & 0.056 & 0.854 & 0.007\\
          $\sigma$ & 0.053 & 0.147 & 0.214 & 0.021 & 0.291 & 0.084 & 0.361 & 0.041\\
         \hline
        \multicolumn{9}{c}{Platt calibrated model} \\
         \hline
          Mean & 0.903 & 0.807 & 0.618 & 0.967 & 0.971 & 0.061 & 0.546 & 0.008\\
          Median & 0.918 & 0.806 & 0.712 & 0.968 & 0.975 & 0.042 & 0.836 & 0.001\\
          $\sigma$ & 0.051 & 0.058 & 0.231 & 0.025 & 0.016 & 0.059 & 0.449 & 0.016\\
         \hline
         \hline
    \end{tabular}
\end{table*}

\subsection{Calibration}
\label{sec:calibration-results}
As mentioned in Sec.~\ref{sec:model_calibration}, calibrating a model is a procedure that can enhance the model predictions to fit the estimated likelihoods. We merged all the data and predictions from the 10 folds into one single dataset, binned the predictions in 10 bins from 0 to 1, and displayed the calibration plots in Fig.~\ref{fig:calibrations}. From such diagrams, we aimed to inspect the model's behavior across the entire range of predictions, and specifically within the extremes (0.0-0.1 and 0.9-1.0).

\begin{table}
    \centering
    \caption{ECE and MCE for uncalibrated, isotonic, and Platt calibrated models.}
    \label{tab:cal-metrics}
    \begin{tabular}{lcccr} 
        \hline  
        Model & Train. ECE & Train. MCE & Val. ECE & Val. MCE\\
         \hline
          Raw & 0.0067 & 0.25 & 0.0069 & 0.28\\
          Isotonic & $7.8\times10^{-6}$ & $8.7\times10^{-6}$ & 0.0023 & 0.21\\
          Platt & 0.0029 & 0.16 & 0.0020 & 0.23\\
         \hline
    \end{tabular}
\end{table}
The uncalibrated model did not show strong deviations from the expected likelihood of predictions (Fig.~\ref{fig:calibrations} left panel) and frequently showed underconfidence (a greater fraction of positives for a given mean predicted value bin), which is a good characteristic of a model intended to avoid false positives. We applied both the isotonic and Platt calibrations, following similar procedures to inspect their effects on the reliability diagram. Any calibration performed on a predictor of any kind needs to be executed using a dataset that has not been used during the training stage. Hence, we calibrated each model using the 10-fold CV datasets, which were created from the original validation datasets. To reduce the overfitting of the calibration, we partitioned each set with an 80-20\% relationship, so we counted both calibration fitting and testing sets. 

\subsection{Results}
\label{sec:results}
\subsubsection{Calibration comparison}\label{sec:cal_compare}
The final median value obtained for the key metric of our 10 cross-validation models was $R@P0.99=0.918\pm0.051$, as reported in Table~\ref{tab:cv-metrics-stats}. However, it is also interesting to assess the entire cross-validation dataset with the 10 models' predictions as a whole and obtain the combined metrics to count with results less sensitive to small validation sets. Measuring the effect of the calibrations on these combined results is also important for comparing different calibration techniques. Hence, we merged the entire dataset with the predictions from each model, both with and without isotonic and Platt calibration, to compare the effect on the disparity of metrics. 

As shown in Fig.~\ref{fig:calibrations} panel left, the model's predictions exhibit minimal deviation from the expected likelihood, suggesting that the uncalibrated output is statistically reliable. The adjustment to the expected fraction of positives fits very well with the expectancy for the dataset, but the predictions deviate significantly within the range from 0.3 to 0.8. The final computations of the numerical calibration metrics are reported in Table~\ref{tab:cal-metrics}. On one hand, the uncalibrated ECE of 0.0067 seemed sufficient to keep the model as is. On the other hand, the MCE of 0.25 is high enough to study the possible enhancements provided by the calibration procedure. Thus, we also analyzed the Platt and Isotonic calibrations to try obtaining more stable metrics.
The isotonic calibrator overfitted the training data (Fig.~\ref{fig:calibrations}, middle panel, and Table~\ref{tab:cal-metrics}), while the validation data was very close to the fit, except for bins placed above the mean predicted value of 0.8. These bins had fewer than 10 samples, and hence, we consider them not very informative and not a good source of information to assume that the isotonic calibrator results are completely overfitted. The Platt calibrator also reported stable results, despite showing strong deviations in the [0-0.1] prediction range for the calibration validation set and excessive underconfidence in the [0.9-1.0] range, as shown in Fig.~\ref{fig:calibrations} right panel. The calibration plots seemed accurate enough for all the cases, and to elaborate the best decision we compared the main metrics of the resulting models after calibration in Table~\ref{tab:cv-metrics-stats}. The Platt calibrator introduces the best results in terms of R@P0.99, with low T@P0.99 uncertainties and the best validation ECE (0.0020, Fig.~\ref{tab:cal-metrics}). However, the isotonic model exhibits more stable metrics at the S@NPV0.999 and especially the S@NPV1.0 thresholds, where only two CV folds fail to reach a value greater than zero. Additionally, the isotonic model also presents the best validation MCE at a value of 0.21.

When a model is ultimately selected for operational use based on a specific evaluation metric, the decision thresholds at which that metric was computed become essential for its practical deployment. This is particularly relevant for probabilistic classifiers, where adjusting the classification threshold directly affects the trade-off between precision and recall. Therefore, a thorough understanding of the model's behavior across different threshold values is crucial to ensure its reliability and effectiveness in real-world applications. As with many probabilistic algorithms, binary classifiers return a floating-point value between 0 and 1, which represents the probability estimated by the model that the input instance belongs to the positive class. Thus, it is essential for the data scientist to select the appropriate thresholds to determine whether a prediction is considered a positive or negative sample. The metrics T@P0.99 and T@P1.0 represent the thresholds applied to the predicted probabilities to obtain precision levels of 0.99 and 1.00, respectively, and are displayed in Table~\ref{tab:cv-metrics-stats}. Isotonic calibration consistently raised both T@NPV0.999 and T@NPV1.0 thresholds. In contrast, Platt calibration showed a slight degradation in T@NPV1.0, while maintaining comparable performance to the uncalibrated model at T@NPV0.999. As NNs are frequently extreme predictors, they report most probabilities as either close to 0 or 1. Therefore, the expected result of model calibration would be the expansion of these extreme predictions across the [0-1] range, as it happened with the isotonic calibration. The precision is a metric based on the predicted positives, and hence, sliding the positive predictions to a wider and lower range would decrease the positive threshold. The original and the Platt calibrated models were unable to reach S@NPV0.999 in 4 of the k-folds. While the isotonic-calibrated model exhibited slightly lower R@P0.99 performance compared to the uncalibrated and Platt-calibrated versions, it consistently delivered more reliable predictions at the negative threshold extremes. Specifically, only 2 out of 10 cross-validation folds failed to reach non-zero values of S@NPV1.0 with isotonic calibration, whereas this occurred in 4 folds for both the original and Platt-calibrated models. These zero-valued metrics indicate that the models were unable to identify any threshold yielding a sufficiently high NPV, effectively rendering the specificity undefined in those cases. Despite the higher variance reported for the isotonic model in some metrics, this is largely due to the greater number of meaningful non-zero values rather than instability. As a result, we selected the isotonic-calibrated model for the remainder of the analysis, as it demonstrated more consistent and usable behavior when assessing low-probability (negative) predictions.

\subsubsection{P@K analytics}
\label{p@k}
Counting with good recall and precision metrics is a promising signal for any kind of model, especially for imbalanced datasets where the positive class is the minority. Furthermore, P@K metrics can provide an additional contribution to the information about the model's behavior on the highest ranked predictions. With such a statistic, we can examine how many negative samples are ranked above the positive ones, thereby uncovering problems hidden by confusion matrix-based metrics like precision and recall, where no assessment of the predicted values is carried out. In Table~\ref{tab:topk-metrics} we report several P@K precision values, together with the number of false positives (FP) found among the K samples set. The P@K500 values only report one FP for the Platt-calibrated model, whereas P@K1000 reports 4 FPs both for the Platt-calibrated and the uncalibrated model. The isotonic model shows significantly better performance, finding its first FP at the 1600th position. The P@K2000 metric shows a marked improvement for the isotonic-calibrated model, achieving a precision of 0.995 and outperforming both the uncalibrated (0.975) and Platt-calibrated (0.993) versions. This improvement highlights the effectiveness of isotonic calibration in refining the model’s ranking capabilities, particularly by widening the output distribution in folds 6 and 7, which originally tended to cluster predictions around 0.5. By redistributing these intermediate values toward more confident scores, the calibration reduced the frequency of ambiguous outputs, thereby lowering both the FPs and FNs rates in high-confidence predictions.

\begin{table*}
    \centering
    \caption{P@K computed for the mixed validation sets with the results of the 10 models combined. Values in parentheses represent the number of false positives that appeared, when any was found. FPs and FNs are reported for a threshold of 0.5.}
    \label{tab:topk-metrics}
    \begin{tabular}{lcccccccr} 
        \hline  
        Model & P@K500 & P@K1000 & P@K1500 & P@K2000 & P@K2200 & FPs & FNs\\
         \hline
          Raw & 1.0 & 0.996 (4) & 0.980 (30) & 0.975 (50) & 0.965 (56) & 78 & 146 \\
          Isotonic & 1.0 & 1.0 & 1.0 & 0.995 (10) & 0.991 (19) & 81 & 126 \\
          Platt & 0.998 (1) & 0.996 (4) & 0.997 (5) & 0.993 (14) & 0.988 (25) & 112 & 108 \\
         \hline
    \end{tabular}
\end{table*}

\subsubsection{Thresholds establishment}
\label{sec:thresholds}
To establish a productive model capable of assessing the reliability of any candidate for a transiting planetary signal, defining proper validation and rejection thresholds is crucial. Based on the analyzed metrics from Table~\ref{tab:cv-metrics-stats}, where we select the isotonic model, we extract a reliable threshold of $0.646\pm0.147$ for a likely planet (LP) signal from the median value of the T@P0.99 metric. This threshold represents the expected prediction value for a false positive probability (FPP) of 0.01.

To consider any result as a likely negative (LN), we need to establish first some metrics for the negative predictions. For this, we used the NPV and specificity. This analysis needs to be performed with a different ratio because the number of negative samples is much higher than positive ones (10 to 1 approximately, as in other studies like ExoMiner). Hence, we selected the thresholds at NPV of 0.999 and computed the sensitivity (Table~\ref{tab:cv-metrics-stats}). The original model contained several folds whose S@NPV0.999 was 0.0 (Table~\ref{tab:cv-metrics}). Despite the isotonic model corrected three of them, leaving only two folds with S@NPV0.999~=~0.0. Under this consideration, we considered that it was still safe to use the T@NPV0.999 median, which is T@NPV999=$0.056\pm0.084$. This is the threshold that we considered for LN classification when using the 10 produced models. 

Once both the LP and LN thresholds are in place, it is also interesting to retrieve thresholds where the model performs perfectly, with neither false positives nor false negatives found. These thresholds are not always possible to encounter or, in case they exist, they might be too low or high to be useful to delimit validation or rejection probabilities. We computed the R@P1.0 and S@NPV1.0 to study both the planet validation and signal rejection boundaries, respectively. Table~\ref{tab:cv-metrics} shows the R@P1.0 obtained for each fold, together with the threshold where the metric was obtained. In general, most of the folds were capable of retaining a high recall for perfect precision. The median threshold for a validated planet (VP) can now be established at a higher or equal probability of $0.961\pm0.021$. In previous studies, such as \cite{exominer2022} and \cite{giacalone2021}, the validation threshold has often been established solely based on achieving adequate precision levels. Moreover, not all studies define a specific rejection threshold. For WATSON-Net, we adopted the same methodology for setting both the rejection and validation thresholds. Consequently, using the S@NPV1.0 values reported in Table~\ref{tab:cv-metrics-stats}, we determined a threshold of $0.007\pm0.041$, which we propose as the validated negative (VN) threshold for our NN.

\begin{figure}
    \centering
    \includegraphics[width=0.7\columnwidth]{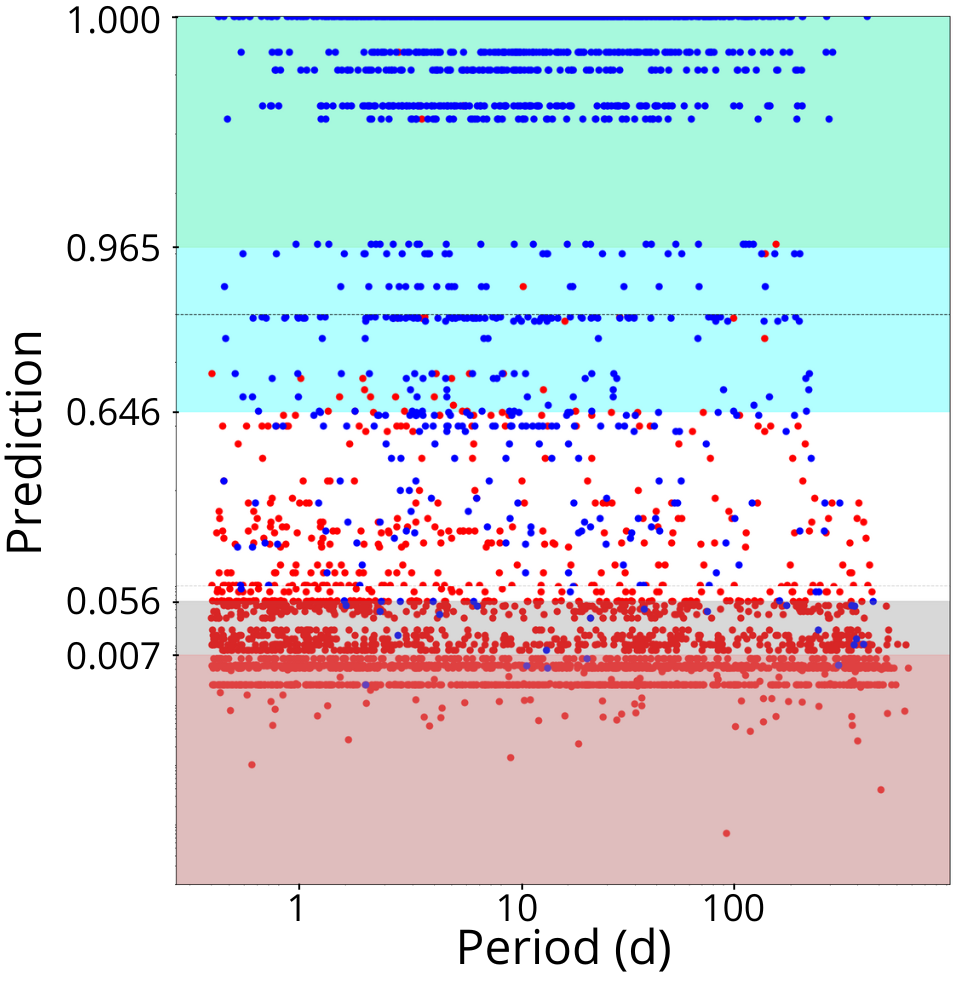}
    \includegraphics[width=0.7\columnwidth]{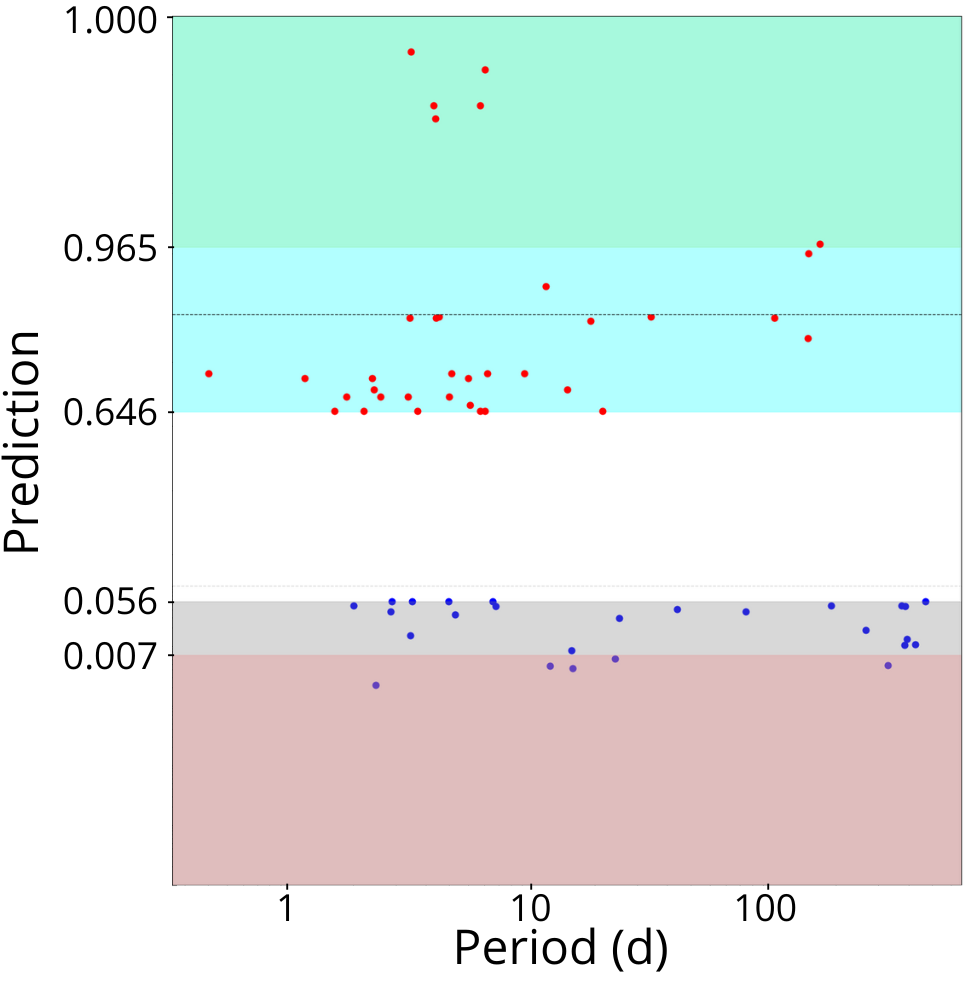}
    \caption{Isotonic model distribution of the validation set predictions for positive (blue) and negative (red) samples. The top light green area matches the validation threshold, and the cyan area matches the likely planet threshold. The gray area matches the likely negative threshold, and the red area matches the validated negative threshold. The top panel displays the entire combined validation dataset, while the bottom panel plots only the FPs above the LP threshold and the FNs below the LN threshold. The horizontal dotted lines at 0.95 and 0.01 represent the change of scale for the top and bottom of the plot, respectively.}
    \label{fig:predictions-distribution}
\end{figure}

Under the proposed median thresholds, we computed the distribution of the predictions in Fig.~\ref{fig:predictions-distribution}. It is essential to note that the thresholds plotted alongside this distribution are computed as the median from each fold value, whereas the plotted predictions are obtained only once from each fold validation set. This is done so because the predictions should not be averaged for the combined dataset: each sample is always used for training, but for one fold, where it is used within the validation set. That is, if we averaged them, we would be using predictions from the training set. More than 91$\%$ of the TPs are properly located within LP boundaries, and more than 84$\%$ lie above the VP region. The majority of negative samples are also located within the LN region, representing more than 97.5$\%$ of the negative signals, and more than 91.5$\%$ of TNs reside within the VN threshold. However, as we set a high standard for rejecting signals, many of the signals remain in the unclear region of predictions ($\sim$9$\%$ of real positives, $\sim$2.5$\%$ of real negatives). Taking into account that the thresholds are plotted based on the median value without their confidence intervals, we could propose widening the LN and LP thresholds up and down to the $median + \sigma$ and $median - \sigma$, respectively, to include most of the negative and positive samples in their proper matching distribution. Despite, a more conservative approach was taken and we chose to keep the thresholds at their obtained median values, as this widening would not only include more accurately classified samples within their appropriate regions, but would also increase the number of incorrectly tagged classes because signals with prediction scores falling between the LP and LN thresholds lie in a region of low statistical confidence. These cases are frequently not suitable for automatic validation or rejection and typically require further observational or contextual follow-up

\subsubsection{False positives}
Based on the LP threshold selected in Sec.~\ref{sec:thresholds}, we inspected the false positives reported by the isotonic model. Fig.~\ref{fig:predictions-distribution} shows that 35 false LPs were found, 6 of them being classified as VPs. We list all these targets together with the likely explanation of their acceptance in Table~\ref{tab:false-positives}. None of the 6 false VPs from the table reported any problematic input for the model, and hence, they could not be labeled as negatives. We can outline the main reasons for negative samples classified as LP in:
\begin{itemize}
    \item Low significance of source offset: The signal source was expected to show deviation from the target, but our computation led to a higher uncertainty than the transit source distance. Therefore, the model is unable to reject a signal based on this metric. The highest prediction for such a target is the KIC 2446113 signal at $P~=~6.72d$ , where the official DVR reported an out-of-transit source offset of 1.232 arcsec ($8.37\sigma$), whilst the WATSON-Net S/N of this value was below $1\sigma$, showing an error higher than the computed value 
    \item Low S/N centroids shift: Both the computed centroids shift in the right ascension and declination did not show the expected deviation from the baseline. The signal with $P~=~4.39d$ for KIC 7732285 showed a centroid shift significance of 2.13 arcsec ($5.13\sigma$) as calculated by the official DVR. However, the both the right ascension and declination centroids offset curves computed by WATSON-Net did not show a clear signal, leading to a misclassification of this TCE with a high value. In this scenario, the official DVR also provided a significant value for the source offset which was also missed by the WATSON-Net data processing. 
    \item No worrisome parameters for rejection: KICs 5544450, 7594098, 6290935, 8261920, 9873759, 12021943, 4067336, and 9770983 did not present any input parameter to make them suspects of a negative classification and did not show noisy data. None of these targets contained problematic metrics that could be spotted by our model given the input data it receives, and hence, their acceptance is explainable.
    \item Low optical ghost significance: The subtraction of the core aperture flux from the halo aperture flux did not produce an expected significant drop in the resulting flux in comparison to the baseline. This frequently happens for very low S/N transits or noisy light curves.
    \item Unclear acceptance: KIC 8681125 (P~=~153.78d) showed a clear feature-less even transit but was not classified as a negative sample. Our hypothesis is that it may be caused by a substantial number of entries within our dataset that are half or double the actual period, which would introduce the same features (either a strong secondary event or odd/even differences). Whenever centroid shifts are spotted, they are often accompanied by optical ghost and/or source offset signatures. This was not the case of KIC 7767559, which showed an important centroid shift, but no other collateral hints were present. The case of KIC 7692093 is clearer, as the only problematic metric was the bootstrap FAP, whose value was high enough for the SPOC to reject the signal. However, it did not seem to cause our model to return a low score on its own. KIC 5632093 is a signal that does not exhibit any problematic metrics in the official Data Validation Report \citep[DVR,][]{dvr1, dvr2}, even when a harmonic TCE is also identified with a Period of 1\,day. KIC 7222086 exhibits a V-shaped transit signature, which is unable to reduce the prediction value despite the rest of the inputs appearing acceptable.
\end{itemize}

With the exception of those cases labeled as unclear acceptances, most misclassifications could be mitigated by refining the data preprocessing pipeline or by incorporating additional input features into the model, particularly for targets affected by significant centroid shifts or source contamination.

\subsubsection{False negatives}
Following the same approach applied to false positives, we conducted a systematic analysis of the false negatives identified by the isotonic-calibrated model, using the LN threshold defined in Sec.~\ref{sec:thresholds}. We illustrate in Fig.~\ref{fig:predictions-distribution} the 26 FNs under the LN area, where 3 of them are classified as VNs. Each of them is listed in Table~\ref{tab:false-positives} together with the likely explanation of their rejection. The 3 candidates lying below the VN threshold do so because their pre-processing was not particularly good, and all of them contained the source offset or a problematic centroids metric in their official DVR as well. Most of the unexpected negatives are correlated to an imprecise data processing, and among the different scenarios, we can divide them into the following categories:

\begin{itemize}
    \item Official signals with problematic metrics: KICs 11446443, 10989274, 1718189, 10937029, 8311864, 11017901, 9141355, and 11138155 are KOIs that contain metrics above the SPOC pipeline thresholds. WATSON-Net agreed with these flags and hence, the confirmed planets were labeled with the negative class.
    \item Differences with official DVRs: KICs 11521793, 9715631, 8410727, 10187017, 3645438, 10397751, 10028792, and 4735826 show significant metrics for rejection, but they were different than those found in their DVRs. It is uncertain why our processing procedure yielded these values compared to the original TCEs.
    \item Imprecise data: KICs 11027624, 10397751, 9141746, 8056665, 8395660, 8891684, and 7100673 presented input information with more noise than expected or improperly normalized, in most cases represented by light curves far from the baseline of 0.5 expected by the model. Although we can not be entirely certain about their rejection, this is probably the main reason behind their low prediction scores.
    \item Unclear rejections: KICs 8873450, 7703955, 6929841. After inspecting the rest of the elements from Table~\ref{tab:false-negatives}, we found several cases where no input data seemed to be significant to cause a rejection. This is the case of KIC 8873450, where an optical ghost effect was reported in the official DVR, but our data did not reveal any outlying value. KIC 7703955 was reported as a very clean candidate from its DVR; however, our resulting data, which also seemed good, led the model to produce a low score.
\end{itemize}

\section{Model Testing}
\subsection{Armstrong et al. (2021) and Valizadegan et al. (2022) validation catalogues}
\begin{figure}
    \centering
    \includegraphics[width=0.7\columnwidth]{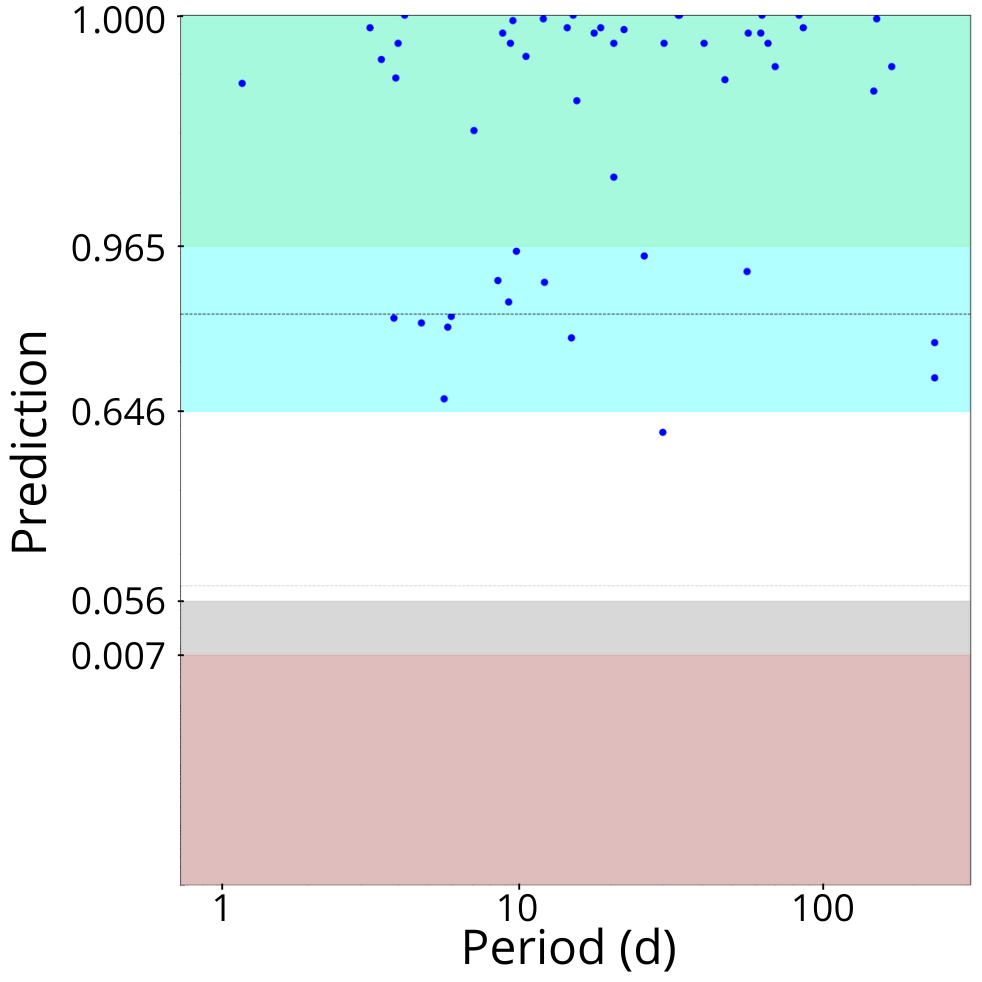}
    \includegraphics[width=0.7\columnwidth]{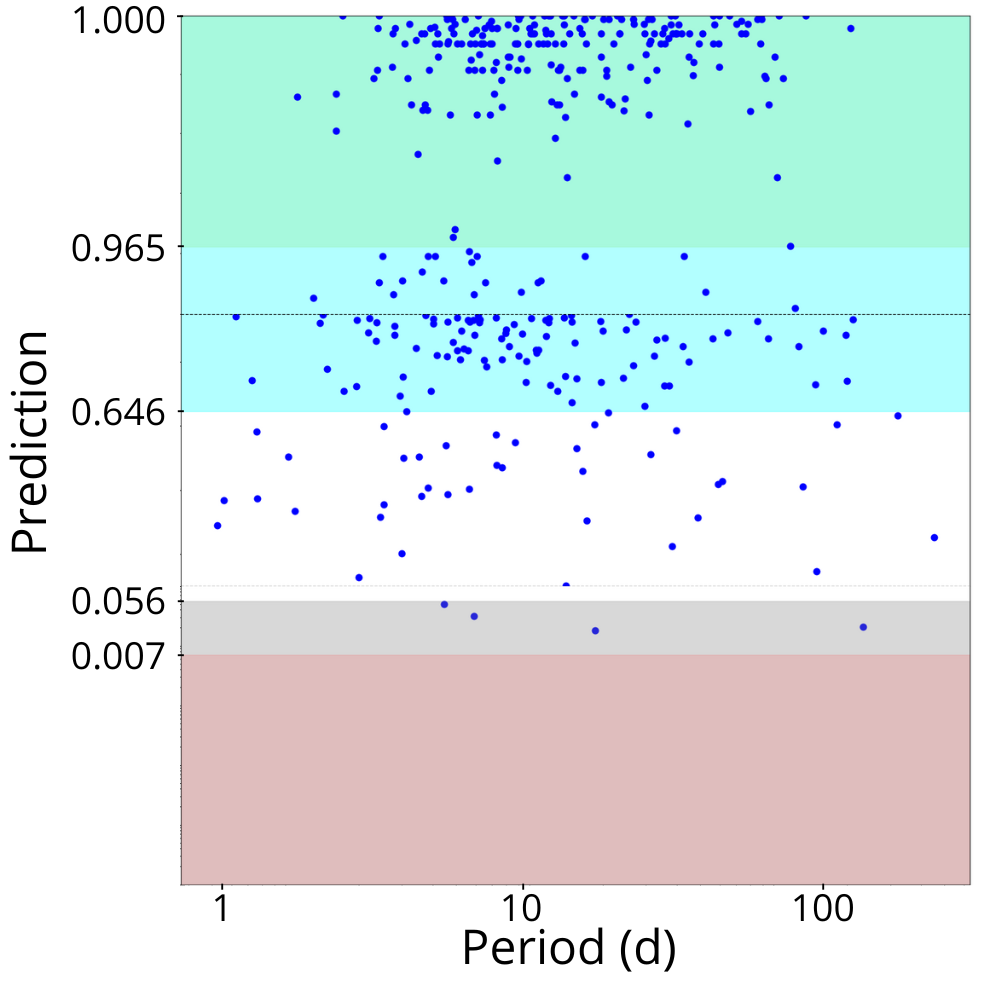}
    \caption{Distribution of predictions for positive (blue) samples for the \protect\cite{2021_gpc}, \protect\cite{exominer2022}, and \protect\cite{exominer2023} test sets. On the one hand, the light-green area corresponds to the validation threshold, and the cyan area corresponds to the likely planet threshold. On the other hand, the gray area corresponds to the likely negative threshold, and the red area to the validated negative threshold. The top panel shows the distribution of predictions for the \protect\cite{2021_gpc} test set, and the bottom panel shows the distribution of predictions for the \protect\cite{exominer2022} and \protect\cite{exominer2023} combined test set. The horizontal dotted lines at 0.95 and 0.01 represent the change of scale for the top and bottom of the plot, respectively.}
    \label{fig:validation-predictions-distribution}
\end{figure}

50 Kepler transiting exoplanet candidates were validated in the \cite{2021_gpc} work. We did not include these confirmed signals in our dataset to avoid using machine-validated data to feed our final model. Consequently, we can also compare the results from our model with this sample set. The only confirmed planet that was below our LP threshold was KIC 8480285 (29.67~d). The complete distribution is plotted in Fig.~\ref{fig:validation-predictions-distribution}. WATSON-Net achieves a recall of 1.0 for a threshold of 0.5, 0.98 for our LP threshold, and a recall of 0.70 when the threshold is set at our VP level.

On the other hand, the foundational paper by \cite{exominer2022} validated 301 planets, followed by \cite{exominer2023}, which validated 69 additional signals, totaling 370 validated exoplanets using ExoMiner. Their validation algorithm consisted of combining the output of the ExoMiner NN with prior probabilities as in \cite{2021_gpc}, considering every signal with a score higher than 0.99 as a validated planet. As we did with \cite{2021_gpc}, we excluded the planets data from our dataset to use it as a test set for our final model. The resulting predictions are shown in Fig.~\ref{fig:validation-predictions-distribution}, where 43 planets are below our LP threshold, from which KICs 10055126 (232.05~d), 3229150 (18.51~d), 3937519 (5.9~d), 4077526 (4.46~d) are located under the LN threshold. KIC 10055126 showed a slightly significant odd/even effect in our data, KIC 3229150 contained a significant source offset metric, KIC 3937519 had a very irregular global view and an imprecise T0, and KIC 4077526 had a very shallow transit depth. In conclusion, all the negatively labeled signals exhibited significant metrics or diagnostic features supporting their rejection as planetary candidates. The recalls for the 0.5, LP, and VP thresholds are 0.92, 0.88, and 0.57, respectively. Therefore, approximately 57$\%$ of the ExoMiner test set falls within our VP region and 88$\%$ within our LP region. The agreement between both NNs is high, but far from a perfect match. One possible explanation lies in the internal use of a physically motivated transit model in the ExoMiner validation framework. This model favors signals that closely resemble idealized planetary transits, potentially boosting the classification scores of certain candidates despite the presence of ambiguous or marginal input data.

Combining both test sets, we obtain a final recall of 0.93, 0.89, and 0.59 for the 0.5, LP, and VP thresholds, respectively. This indicates that the scores are dominated by the ExoMiner test set, as it contains 7.5 times more samples.

\subsection{TESS Confirmed Planets \& False Positives}
\begin{figure}
    \centering
    \includegraphics[width=0.7\columnwidth]{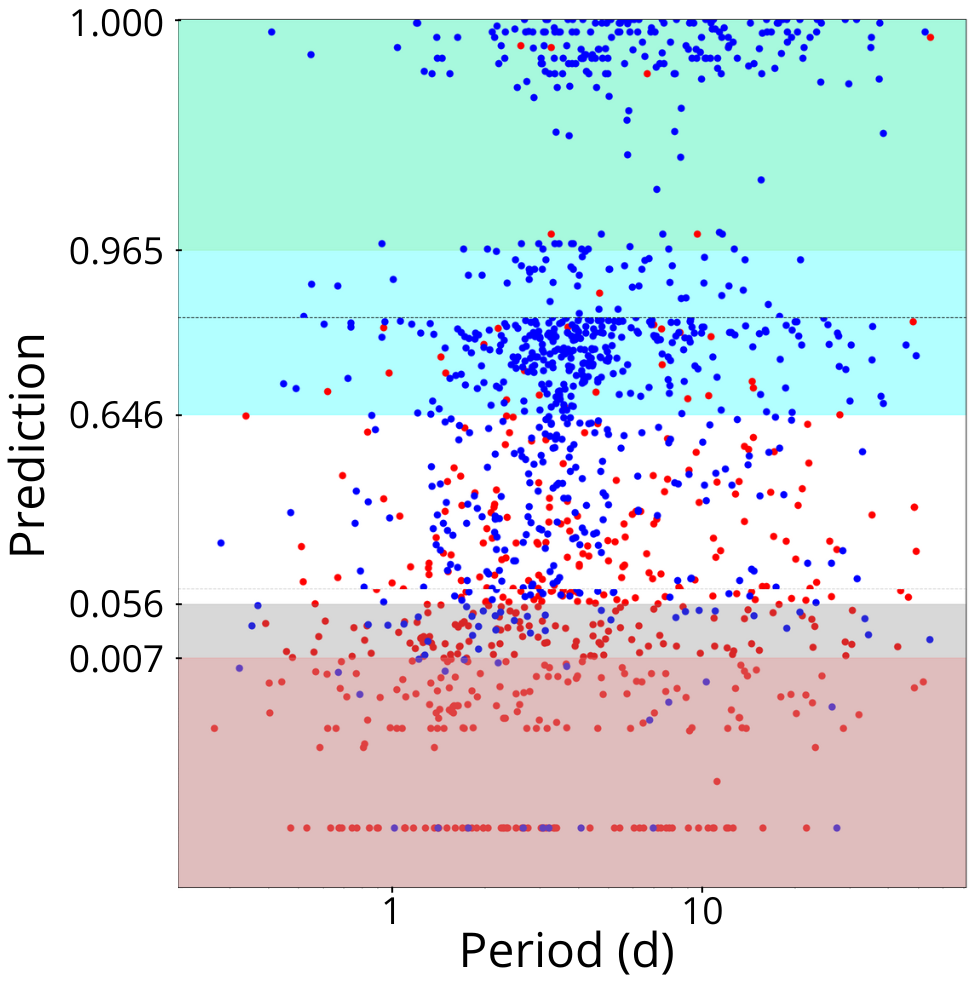}
    \includegraphics[width=0.7\columnwidth]{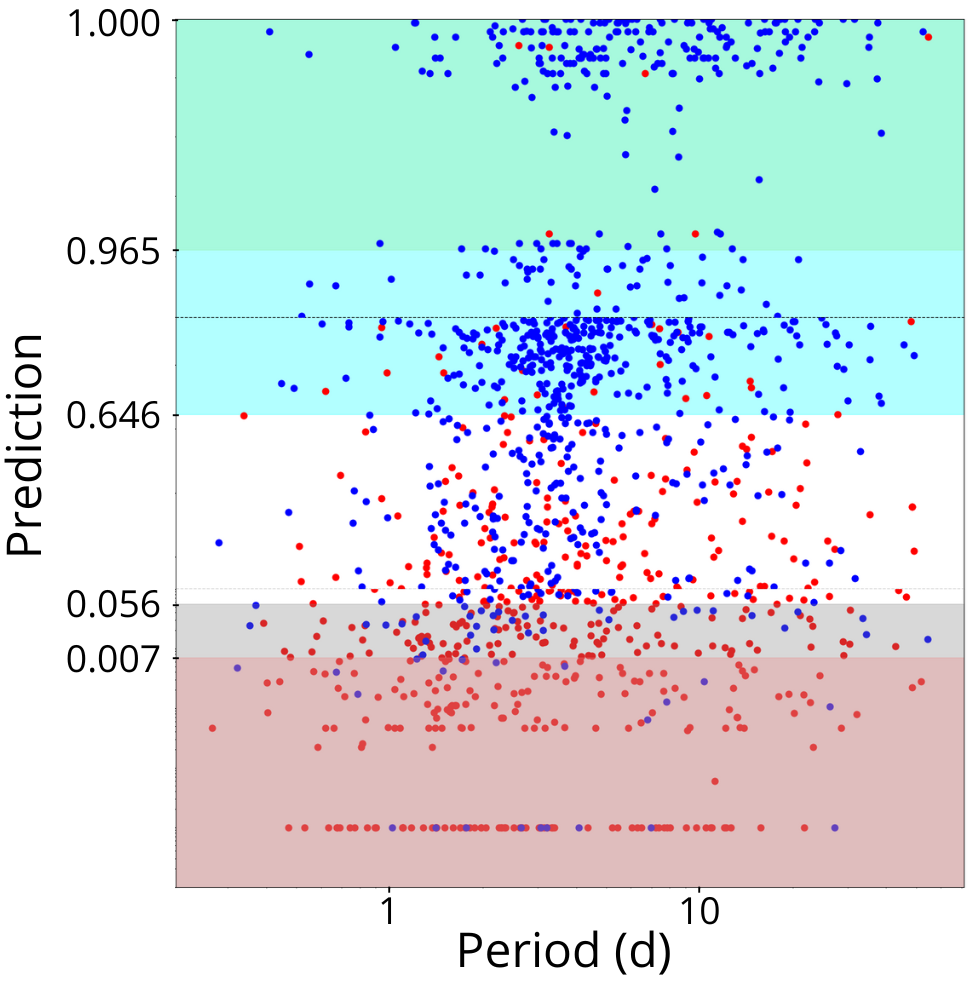}
    \caption{Distribution of predictions for positive (blue) and negative (red) samples. The light-green area corresponds to the validation threshold, and the cyan area corresponds to the likely planet threshold. The gray area corresponds to the likely negative threshold, and the red area to the validated negative threshold. The top panel shows the distribution of all the predictions for the TESS test set. The bottom panel displays the distribution of incorrect classifications, including false positives (FPs) above the LP threshold and false negatives (FNs) below the LN threshold. The horizontal dotted lines at 0.95 and 0.01 represent the change of scale for the top and bottom of the plot, respectively.}
    \label{fig:tess-predictions-distribution}
\end{figure}

We compiled a reference set of confirmed planets (CPs) and validated false positives (FPs) from the TESS Objects of Interest (TOIs) catalog available on the ExoFOP-TESS platform\footnote{\url{https://exofop.ipac.caltech.edu/tess/view\_toi.php}}. Hence, the classification of each target as a CP or FP was initially based on the ``TESS Disposition'' field provided in that catalog. From this list, our pipeline presented a few sporadic failures at the data preparation stage for some targets, which were discarded and will be analyzed in the future. Our final TESS test set consisted of 1310 signals, from which 806 were CPs or known planets (KPs) and 504 FPs, all of which had an officially assigned TOI. The same data preparation pipeline was used to produce the inputs for each signal, where we only used the TESS sectors data that the TOIs were alerted with. The final predictions are plotted in Fig.~\ref{fig:tess-predictions-distribution}. We used the isotonic model as we did in the rest of the work, obtaining a precision of 0.92, a recall of 0.76, and an accuracy of 0.81 for the standard threshold of 0.5. The precision rises to 0.95 and the recall falls to 0.69 when using the LP threshold. Finally, choosing the VP threshold results in a precision and recall of 0.97 and 0.26, respectively. The model achieved precisions of 0.99 and 1.0, reporting both R@P0.99 and R@P1.0 of 0.10 at a threshold of 0.997. However, it was impossible to achieve high NPV values, and therefore, no thresholds for rejection could be established. This aligns with the high uncertainty of the obtained T@NPV0.999 and T@NPV1.0 values for the Q1-Q17 validation dataset. The FPs above the LP threshold are listed in Table~\ref{tab:tess-false-positives}, where the explanation for each acceptance is provided.

The detected number of FNs is larger than the FPs for the TESS test set. This aligns with the same behaviour found with the validation dataset. As we prioritized precision over recall to select our best model, we would expect to have more reliable predictions for the positive class than those for the negative class. All the FNs whose predictions are within the LN or VN region are described in Table~\ref{tab:tess-false-negatives}.

As an experiment, we attempted to predict the same test set using the Platt-calibrated model, achieving a precision of 0.93 and a recall of 0.77, with R@P0.99 and R@P1.0 of 0.12 at the 0.5 threshold. That is a relatively significant improvement, but it is still close to the isotonic results. Most of the FNs predicted by WATSON-Net are hot Jupiters (P~<~10d, R~>~6$R_\odot$), which tend to show blended light curves, secondary transits, or centroid shifts in a number of cases. When excluding these types of signals from the TESS test set, 836 signals remain in the test set (332 CPs and 504 FPs). WATSON-Net reports a precision of 0.83, a recall of 0.80, and an accuracy of 0.86. Hence, the recall is improved at the cost of reducing the precision, significantly increasing the accuracy of the model for this set. This means that hot Jupiters are more difficult to classify and are often misidentified as eclipsing binaries. This challenge is not specific for Watson-NET, but it is also observed in other works like Exominer, where in Sec~6.2 \citep{exominer2022} the assessment of metrics necessary to identify these kind of planets is not properly understood by the model.

\subsection{Metrics comparison}
% All the metrics presented in this work can be better understood when compared to the current state of the art. As of today, ExoMiner is the most reliable machine-based binary predictor for the Kepler Q1-Q17 dataset, as it offers top-level precision values while maintaining a high recall, achieving an R@P0.99 of 0.936 with its CV-trained models. We have not reached the same top bar for this metric, obtaining R@P0.99~=~0.918$\pm$0.051 for our raw and Platt-calibrated models, and R@P0.99~=~0.903$\pm$~0.053 for the isotonic model (see Table~\ref{tab:cv-metrics-stats}). Hence, WATSON-Net ranks second (see Table~\ref{tab:ranking}), clearly outperforming the GPC model from \cite{2021_gpc}, which achieved R@P0.99~=~0.761. However, while ExoMiner reports a higher nominal value of R@P0.99 for Kepler data, the difference with respect to WATSON-Net lies within one standard deviation, suggesting that both performances could be statistically comparable at the 1$\sigma$ level. No data related to sensitivity, NPV, or R@P1.0 were provided in other works, and therefore, no further analysis can be done in relation to them. 

The metrics presented in this study are better contextualized through comparison with established methods. ExoMiner currently represents the most reliable machine-based binary predictor for the Kepler Q1–Q17 dataset, demonstrating high precision and recall, and achieving an R@P0.99 of 0.936 with cross-validation (CV)-trained models. The performance of WATSON-Net falls within one standard deviation of ExoMiner’s reported value, with R@P0.99 = 0.918$\pm$0.051 for the raw and Platt-calibrated models, and R@P0.99 = 0.903$\pm$0.053 for the isotonic model (see Table~\ref{tab:cv-metrics-stats}). These results indicate that both methods are statistically comparable at the 1$\sigma$ level. Hence, WATSON-Net ranks second overall and outperforms other available models (see Table~\ref{tab:ranking}).

% \begin{table}
%     \centering
%     \caption{Ranking of classifiers by R@P0.99 values, based on Table~5 from \protect\cite{exominer2022}}
%     \label{tab:ranking}
%     \begin{tabular}{lccr} 
%         \hline  
%         Method & R@P0.99 & R@P0.975 & R@0.95\\
%          \hline
%         Robovetter $^1$ & 0.0 & 0.0 & 0.976 \\
%         AstroNet $^2$ & $0.198\pm0.174$ & $0.343\pm0.143$ & $0.642\pm0.012$ \\
%         ExoNet $^3$ & $0.479\pm0.190$ & $0.643\pm0.090$ & $0.806\pm0.001$ \\
%         % RFC $^4$ & $0.563\pm0.314$ & $0.776\pm0.115$ & $0.935\pm0.001$ \\
%         GPC $^4$ & $0.716\pm0.156$ & $0.848\pm0.101$ & $0.940\pm0.001$ \\
%         Watson-NET $^5$ & $0.903\pm0.053$ & $0.919\pm0.051$ & $0.952\pm0.052$ \\
%         ExoMiner $^6$ & $0.939\pm0.030$ & $0.964\pm0.009$ & $0.988\pm0.000$ \\
%         \hline
%     \end{tabular}
%     \tablefoot{1.~\cite{robovetter2016}, 2.~\cite{astronet2019}, 3.~\cite{exonet}, 4.~\cite{2021_gpc}, 5.~This work Sec.~\ref{sec:model}, 6.~\cite{exominer2022}.}
% \end{table}

\begin{table}
    \centering
    \caption{Ranking of classifiers by R@P0.99 values, based on Table~5 from \protect\cite{exominer2022}}
    \label{tab:ranking}
    \begin{tabular}{lccc} 
        \hline  
        Method & R@P0.99 & R@P0.975 & R@0.95 \\
        \hline
        Robovetter $^1$ & \makecell[c]{0.0} & \makecell[c]{0.0} & \makecell[c]{0.976} \\

        AstroNet $^2$ & \makecell[c]{0.198 \\ $\pm$0.174} & \makecell[c]{0.343 \\ $\pm$0.143} & \makecell[c]{0.642 \\ $\pm$0.012} \\

        ExoNet $^3$ & \makecell[c]{0.479 \\ $\pm$0.190} & \makecell[c]{0.643 \\ $\pm$0.090} & \makecell[c]{0.806 \\ $\pm$0.001} \\

        GPC $^4$ & \makecell[c]{0.716 \\ $\pm$0.156} & \makecell[c]{0.848 \\ $\pm$0.101} & \makecell[c]{0.940 \\ $\pm$0.001} \\

        Watson-NET $^5$ & \makecell[c]{0.903 \\ $\pm$0.053} & \makecell[c]{0.919 \\ $\pm$0.051} & \makecell[c]{0.952 \\ $\pm$0.052} \\

        ExoMiner $^6$ & \makecell[c]{0.939 \\ $\pm$0.030} & \makecell[c]{0.964 \\ $\pm$0.009} & \makecell[c]{0.988 \\ $\pm$0.000} \\
        \hline
    \end{tabular}
    \tablefoot{1.~\cite{robovetter2016}, 2.~\cite{astronet2019}, 3.~\cite{exonet}, 4.~\cite{2021_gpc}, 5.~This work Sec.~\ref{sec:model}, 6.~\cite{exominer2022}.}
\end{table}

To facilitate a consistent comparison with previous studies, the P@K metrics were computed over the aggregated validation sets, as detailed in Sec.~\ref{p@k} and summarized in Table~\ref{tab:topk-metrics}. The uncalibrated and Platt-calibrated models achieved P@K1000 values comparable to that of the GPC model (0.996), and only marginally below the performance of ExoMiner (0.999). In contrast, the isotonic-calibrated version of WATSON-Net, which was ultimately selected for production, maintained a perfect score of P@K1500=1.0 and an outstanding P@K2200=0.991. These results surpass those of ExoMiner (0.985) and GPC (0.954) at the same K value, indicating that the isotonic-calibrated model is more effective at ranking true positives at the top of its prediction list. This suggests a slightly higher reliability when interpreting high-confidence outputs from WATSON-Net.

\begin{figure}
    \centering
    \includegraphics[width=0.99\linewidth]{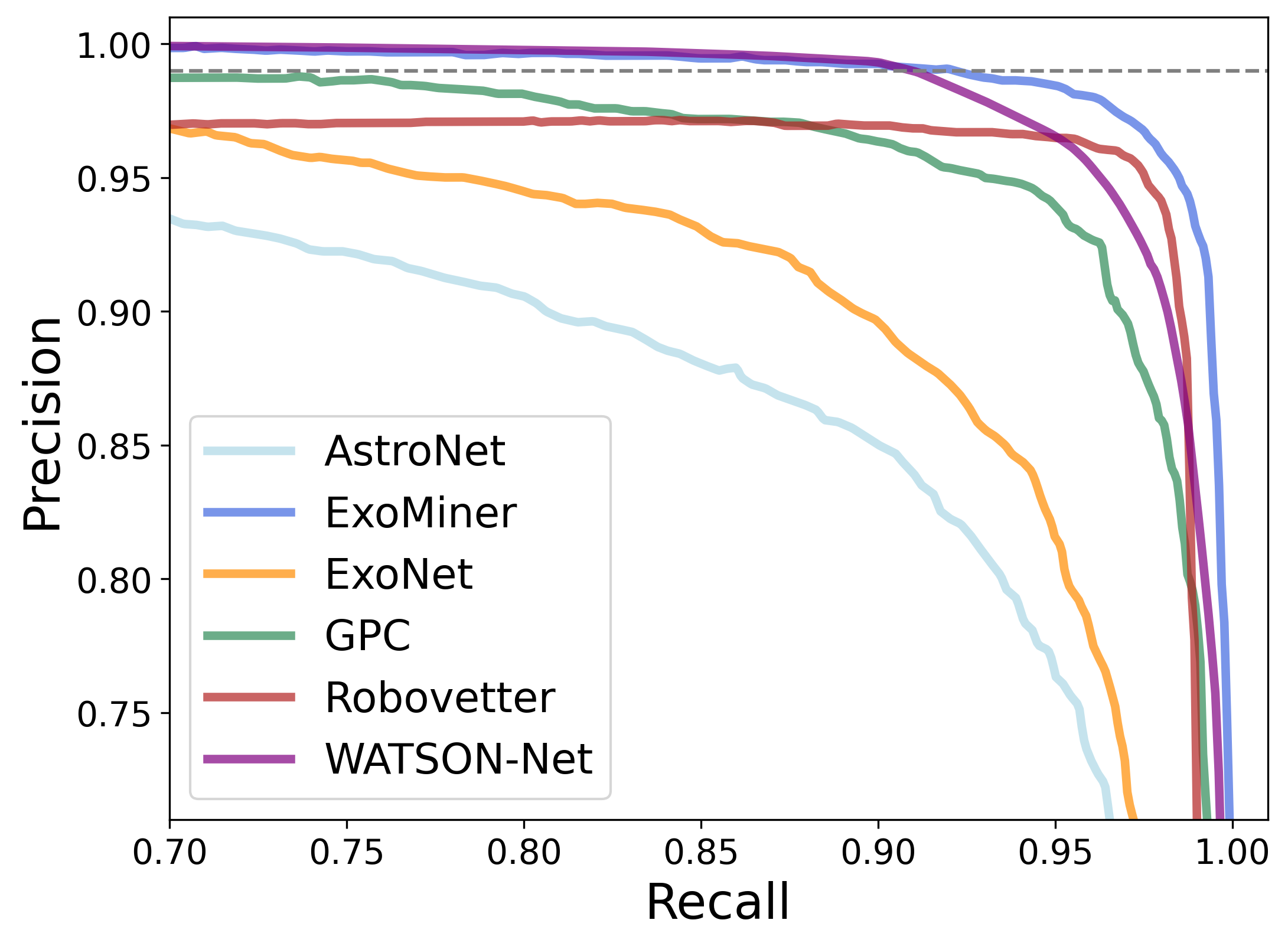}
    \caption{Precision-Recall curve for all the reported models in Table~\ref{tab:ranking}, based on the Fig.~11.a from \protect\cite{exominer2022} and including WATSON-Net results.}
    \label{fig:prauc}
\end{figure}

An effective framework to assess the comparative reliability of classification models in identifying true positives is provided by the Precision–Recall (PR) curve. This representation illustrates the trade-off between precision and recall across a range of classification thresholds, thereby offering a robust diagnostic of model behavior in the presence of class imbalance. In Fig.~\ref{fig:prauc}, we reproduce and extend Fig.~11.a from \citet{exominer2022} by incorporating the PR curve of WATSON-Net computed using the validation set. The results indicate that WATSON-Net consistently achieves slightly higher precision than ExoMiner for recall values below 0.9 and precision values higher than 0.995, and overall attains an Area Under the Curve (AUC) of 0.993. This performance positions WATSON-Net as the second-best model in terms of AUC, surpassing GPC (0.982) and closely approaching ExoMiner (0.995).

The ExoMiner neural network demonstrated excellent calibration performance, with ECE = 0.001 and MCE = 0.067, whereas the GPC model yielded ECE = 0.009 and MCE = 0.145. The Platt-calibrated model returned ECE = 0.002, whilst our finally selected isotonic model yielded ECE = 0.0023 and MCE = 0.21. Although the MCE value is higher, this deviation occurred in the bin with the smallest number of samples and was therefore deemed statistically insignificant. For this reason, ECE was prioritized as the best metric for comparison. Based on this criterion, the WATSON-Net isotonic model demonstrated the second-best calibration performance, only behind ExoMiner.

\cite{exominer2022} reported the results of their simplified model with a subset of TESS TOIs that were either confirmed or rejected. As ExoMiner was fed with metrics that are only publicly available for the Kepler Q1–Q17 catalog, its authors did not compute the same metrics for the TOIs. A more basic version was trained in the same work without using single real numbers. That is, only time series were used for their TESS model, and they named it ExoMiner-Basic. With this model, \cite{exominer2022} achieved a precision of 0.88 and a recall of 0.73, but neither R@P nor S@NPV metrics were reported. As part of the WATSON-Net development, a data preparation pipeline was constructed to generate the same metrics for both Kepler and TESS data. As a consequence, testing on the TOIs test set could be carried out using the final isotonic-calibrated model, which includes all the single-value metrics that ExoMiner-Basic could not incorporate in their tests. Acknowledging this methodological difference, WATSON-Net outperformed ExoMiner-Basic with a precision of 0.92 and a recall of 0.76. Additionally, R@P0.99 and R@P1.0 values of 0.1 were reached, suggesting the possibility of establishing a specific LP or VP threshold for the TESS targets, without further fine-tuning, at a value of 0.997. However, a sufficiently high NPV value could not be achieved, thus preventing the definition of LN or VN thresholds. Subsequently, \cite{exominer2025} introduced an updated version of ExoMiner trained jointly on Kepler and TESS data. This new approach combines both time series and single-value metrics from the two missions, resulting in significant performance improvements for TESS targets. For their TESS dataset, which includes a larger sample of negative examples, they reported a precision of 0.926 and a recall of 0.945. Although direct comparison is limited due to the large dataset differences, these results clearly illustrate that incorporating mission-specific data during training is crucial to building more robust classifiers. 

\subsection{Model Explainability}
Machine-based predictors and classifiers are powerful tools to enhance procedures where prioritization or strategic decision-making is necessary. Under these circumstances, it is essential for these algorithms to provide as much information as possible about their reasoning behind their results. This concept is called explainability \citep{explainability}, and it is a fundamental consideration when models are deployed in production. Some traditional ML predictors can provide easy explainability from their bedrock: linear regression directly relates its coefficients with the predicted features, decision trees split rules are easily understandable, or logistic regression feature weights show logarithmic odds changes for binary classification. Other approaches require supportive methods to explain their behavior: SHAP \citep{explainability_shap} provides general and local explainability for any machine learning model, while LIME \citep{explainability_lime} is capable of incorporating some level of explainability for text or image data. However, these tools are insufficient to provide a comprehensive understanding of the decisions made by complex models and are sometimes tailored for specific cases, as is the case with TreeSHAP \citep{explainability_treeshap}, a customized version of SHAP that specializes in tree-based models. Some of these methods can be used with NNs but with poor results, and hence other strategies are proposed for different NN architectures to provide better insights on their predictions, as it happened with Grad-CAM \citep{explainability_gradcam}, which can be used to inspect the relevant regions of the features extracted by a CNN.

In this work, an ensemble of different CNN branches has been created, incorporating dense layers; consequently, no specific adaptations for CNNs are required. This approach is very similar to the one taken by \cite{exominer2022} for ExoMiner, where the introduced explainability framework was based on specific single-branch occlusions to study their effect on the original prediction. This idea is very promising and reasonable for a multi-branch model where each input can be easily nullified. Nonetheless, nulling specific inputs can be dangerous if the model was not trained with a substantial amount of scenarios where this situation was reproduced in the training set. This is something that was not mentioned in \cite{exominer2022}. Even when their explainability framework seems solid, it could produce unreliable outputs in certain untested scenarios due to the presence of extreme input values (zeros). To mitigate this effect and make the model aware of zeroed inputs on each branch, the training procedure of WATSON-Net included the Branch Dropout concept introduced in Sec.~\ref{sec:nn_design}. In addition to the branch nullification, WATSON-Net also includes the variation from one extreme to the other of single number inputs: period, duration, planetary radius, number of transits with enough measurements, number of good transits out of the total number of transits, secondary event albedo statistic, secondary event temperature statistic, odd-even factor, bootstrap FAP, stellar radius and stellar temperature. This approach would help to understand the impact of significant changes on these values on the model predictions. A total of 44 different scenarios are explored, comprising 8 branch occlusions and 36 single-value switches. To test the WATSON-Net explainability support, all the FPs above the VP threshold (Table~\ref{tab:tess-false-positives}) and the FNs from the TESS test set within the VN threshold (Table~\ref{tab:tess-false-negatives}) were explored. 

Firstly, we explored the framework for the \tess FPs above the VP threshold. The results for TICs 300013921 (P~=~54.66d), 301051430 (P~=~3.27d), and 230017325 (P~=~9.69d) showed that the even and odd branches provide the largest boost to the prediction, closely followed by a large source offset increment, the optical ghost branch, and the main transit view branch. This means that there are many features that create large output weights, which are not neutralized by the significance of the centroids' branches. The cases of TIC 97158538 (P~=~2.60d), 82707763 (P~=~6.66d), and 470974162 (P~=~3.26d) have more complexity because the inputs contained no problematic metrics. Explaining these targets shows that the most effective measurement to reduce the prediction is the source offset increment, except for TIC 470974162, where the even and optical ghost branches contribute each significantly to $\sim0.30$ of the prediction score. This suggests that these two branches may be the most relevant for the model decision in this specific case.

We carried out the same analysis for FNs under the VN threshold. The understanding of the FNs scenarios is more complex because extreme negative classifications provide very low scores, where each branch contributes with tiny values. Therefore, branch occlusion does not provide any score variation in many cases, and switching single values is often insufficient to improve the predictions significantly if not done together with other inputs. The explanation of TIC 267574918 (P~=~1.41d) did not show any variation under any of the combinations because there is more than one strong negative indicator. TIC 164786087 (P~=~4.08d), 380619414 (P~=~2.66d), 158324245 (P~=~1.76d), and 138359318 (P~=~1.02d) are interesting cases because they showed two problematic parameters: the stellar radius (3$R_\odot$) and extreme planetary radii, which were derived from the transit depths and stellar radius. Consequently, the only variation that provoked a significant change in the prediction was the switch of the planet radius to 0.1$R_\oplus$, which increased the prediction from 0.000010 to 0.970648, 0.516017, 0.121363, and 0.109449, respectively, where the last three showed different levels of centroids shift significance, leading to their lower score increases. The predictions for TIC 329148988 (P~=~27.27d), 363548415 (P~=~6.77d), and 142276270 (P~=~26.32d) do not improve under any explored scenario, likely due to the lack of valuable information provided by any of the transit views. TIC 404243877 (P~=~3.20d), 164173105 (P~=~3.07d), 35516889 (P~=~0.79d), 27491137 (P~=~10.36d), 134537478 (P~=~1.71) contain several branches or single values leading to a negative classification; thus, no scenario produces a remarkable increase in their score. The TIC 392476080 (P~=~0.67d), 424865156 (P~=~2.20d), and 129979528 (P~=~1.22) scores were improved up to 0.10 when the secondary transit or the global view was occluded, but as they also contain other reasons for rejection (e.g., large planetary radius), the prediction can not be improved more. TIC 16740101 (P~=~0.32d) input branches were computed using a longer transit duration than the real one, and thus, the explainability scores can not increase. TIC 37718056 (P~=~3.66) data was computed using a wrong transit epoch, which caused improper inputs, and we did not explore it.

Overall, the analysis of the misclassified cases indicates that most WATSON-Net FPs and FNs arise when our locally pre-processed diagnostic inputs exhibit low significance, whereas the corresponding quantities appear clearly significant in the official DVRs. This systematic discrepancy suggests that the majority of misclassifications originate from differences in input generation, rather than from the neural network architecture itself.

\subsection{Integration into The SHERLOCK PIPEline}
The practical use of WATSON-Net as a vetting engine requires seamless integration with existing analysis frameworks employed by the community to validate and prioritize exoplanet candidates. For this purpose, we implemented WATSON-Net also as a dedicated module within the SHERLOCK pipeline \citep{devora2024}, enabling its direct application to any signal processed through this end-to-end detection and vetting framework. 

SHERLOCK is an open-source, modular pipeline designed to perform exhaustive transit searches and vetting procedures on light curves from Kepler, K2, and TESS. Its architecture comprises successive modules that preprocess light curves, detect transiting signals, compute diagnostic metrics, and produce detailed vetting reports for each candidate. Before WATSON-Net, SHERLOCK included a preliminary heuristic-based module named WATSON, which computed specific statistical features and metrics to aid human vetting decisions. However, the process remained semi-automated and required expert interpretation. With the integration of WATSON-Net, SHERLOCK gains a fully automated and statistically interpretable vetting engine capable of classifying candidates as likely planets or false positives based on a complex set of inputs and learned features from training on Kepler DR25 and TESS datasets. The module is directly accessible containing both the model and its data preparation submodules. 

The integration is designed to be modular and mission-agnostic, supporting Kepler, K2, and TESS targets as long as the required data products are available. In case any of the input branches cannot be generated (e.g. due to missing centroid data), the Branch Dropout mechanism ensures that predictions remain stable without raising execution errors, albeit with reduced interpretability. Upon execution, the WATSON-Net module appends to each candidate’s vetting report: 1) The predicted probability of planetary nature (between 0 and 1), 2) The explanation of the label assignment, including the relevant threshold used, and the explored explainability scenarios presented in this work, 3) diagnostic plots of the input data used by the neural network for expert inspection if required. These outputs are fully compatible with SHERLOCK’s report generation and can be incorporated into automated follow-up prioritization workflows or planet validation pipelines.

The inclusion of WATSON-Net within SHERLOCK significantly enhances its vetting capabilities by introducing state-of-the-art deep learning classification based on extensive training on Kepler and TESS data. This integration transforms SHERLOCK from a semi-automated detection pipeline into a full open-sourced and community-driven detection-to-vetting framework, capable of outputting planet candidate catalogs with minimal manual intervention.

\section{Conclusions}
In this study, we present WATSON-Net, a novel deep-learning architecture developed for the vetting and validation of transiting exoplanet signals detected by space-based missions such as Kepler and TESS. The model is designed as a binary classifier to accurately discriminate between genuine planetary transits and a wide variety of astrophysical and instrumental false positives. It leverages a comprehensive set of input features, including phase-folded light curves, centroid shifts, optical ghost diagnostics, and both stellar and transit-specific parameters. Our approach places particular emphasis on model explainability, reproducibility, and accessibility.

The model was trained using a rigorous 10-fold cross-validation scheme on the Kepler DR25 dataset and evaluated on both internal validation sets and independent test samples. WATSON-Net achieved a median recall at 99\% precision (R@P0.99) of 0.903, ranking as the second-highest-performing model for Kepler vetting tasks, surpassed only by ExoMiner. Furthermore, it outperforms all previously published machine learning classifiers that are not specifically fine-tuned to TESS data. When applied to TESS targets, WATSON-Net exhibited strong generalization capabilities, becoming the most accurate non-fine-tuned vetting model evaluated to date. The recently introduced ExoMiner++, fine-tuned on both Kepler and TESS data, achieves state-of-the-art performance on TESS targets. Although not directly comparable to our framework, since its training set includes a larger fraction of obvious negatives, it illustrates the gains achievable through fine-tuning. In contrast, WATSON-Net already provides competitive results on TESS without any mission-specific retraining, reinforcing its role as a robust and broadly applicable open-source tool. Looking ahead, developing a fine-tuned version trained on mission-specific data will be a key step to further enhance the reliability of WATSON-Net predictions.

A remarkable difference between Exominer and WATSON-Net lies at the data processing stage. We computed our own centroids and optical ghost light curves, together to our own odd/even, albedo and temperature statistics, while Exominer used the official metrics computed by the SPOC. That creates more difficulties for WATSON-Net to reach the same R@P as data processing differences are expected to be frequent and hence, matching the official labels becomes slightly harder. 

A comprehensive calibration analysis, employing both isotonic regression and Platt scaling, revealed that the uncalibrated model predictions already showed a strong alignment with empirical likelihoods. While both techniques yielded modest improvements in standard calibration metrics, the isotonic-calibrated version was ultimately selected for production due to its robustness across probability thresholds and superior specificity under stringent classification criteria. Based on these results, we defined interpretable and operational prediction thresholds corresponding to four confidence-based labels: Likely Planet (LP), Likely Negative (LN), Validated Planet (VP), and Validated Negative (VN). This classification scheme provides a practical and consistent framework for interpreting model predictions in downstream vetting applications. However, these limits can only be applied to Kepler targets. When extrapolating these thresholds to TESS data, we found that the model maintained high precision and sufficient recall for positive classifications, enabling the possibility of establishing VP and LP tags. However, it failed to achieve sufficient reliability in the negative regime, which prevented the robust definition of different LN and VN thresholds.

A detailed analysis of false positive and false negative cases, with the useful help of the developed explainability framework, revealed that most misclassifications are originated from input features with low signal-to-noise ratios or from residual artifacts introduced during the preprocessing stage. These results highlight the crucial role of high-quality data curation and suggest potential improvements in the computation of centroid shifts and optical ghost diagnostics. Nevertheless, the consistently low misclassification rates, together with the strong performance of the model in precision-at-K evaluations and its stability across decision thresholds, highlight its readiness for integration into scientific vetting pipelines.

Ultimately, WATSON-Net represents a substantial advancement in the automated vetting of transiting exoplanet candidates, offering a high-performing, transparent, partially self-explainable and versatile framework applicable to both current and upcoming space missions. Rather than being designed to generate large-scale catalogs of validated planets, WATSON-Net has been conceived as a scientific tool to assist and empower researchers in their own vetting workflows. Its integration into the SHERLOCK pipeline reflects this philosophy, providing users with an interpretable probabilistic score that can guide the prioritization and assessment of individual signals within user-defined surveys. All code, data, and calibration procedures are made publicly available through the \dearwatson package, ensuring full reproducibility and fostering broader adoption across the community. Future work will focus on refining its applicability across missions such as TESS and PLATO, as well as exploring its potential in multi-mission vetting contexts.

% \section{Future work}

% \subsection{Inclusion of physical scenarios}

% \subsection{Model for each branch}

% \subsection{TESS fine-tuning}

% \textcolor{red}{Explainability framework needs further work to explain combinations of values or branch views}

% \subsection{Multi-tag model}

% \subsection{New architectures}

% \textcolor{red}{MIGRATE TO TRANSFORMERS ARCHITECTURE}

% \subsection{PLATO adaptations?}

\begin{acknowledgements}
We thank the anonymous referee for the helpful comments. F.J.P. acknowledges financial support from the Severo Ochoa grant CEX2021-001131-S funded by MCIN/AEI/10.13039/501100011033 and through the project PID2022-137241NB-C43. 
We acknowledge support from the Spanish Ministry of Science through the project PID2019-107061GB-C64/SRA (State Research Agency/10.13039/501100011033). Funding for open access charge: Universidad de Granada / CBUA. J.C.S. acknowledges support from the project PID2023-149439NB-C43 funded by MICIU/AEI/10.13039/501100011033 and by “ERDF A way of making Europe”.
\end{acknowledgements}

\bibliographystyle{aa}
\bibliography{biblio}
\begin{appendix}

\section{Neural network design: additional data}
% %%{ init: {
%     'flowchart': { 'curve': 'stepBefore' },
%     'theme': 'base',
%     'themeVariables':
%         { 'fontSize': '30px', 'fontFamily': 'Inter'}
% } }%%
% flowchart TD
%     A[Global flux 300x1] --> B[Gaussian Noise]
%     B[Gaussian Noise] --> C[Conv1D f=16 ks=90]
%     C[Conv1D f=16 ks=90] --> D[MaxPooling1D ps=10 s=1]
%     D[MaxPooling1D ps=10 s=1] --> E[SpatialDropout1D]
%     E[SpatialDropout1D] --> F[Conv1D f=32 ks=30]
%     F[Conv1D f=32 ks=30] --> G[MaxPooling1D ps=4 s=1]
%     G[MaxPooling1D ps=4 s=1] --> H[SpatialDropout1D]
%     H[SpatialDropout1D] --> I[Conv1D f=64 ks=15]
%     I[Conv1D f=64 ks=15] --> J[MaxPooling1D ps=3 s=2]
%     J[MaxPooling1D ps=3 s=2] --> K[SpatialDropout1D]
%     K[SpatialDropout1D] --> L[Conv1D f=128 ks=8]
%     L[Conv1D f=128 ks=8] --> M[MaxPooling1D ps=2 s=2]
%     M[MaxPooling1D ps=3 s=2] --> N[SpatialDropout1D]
%     N[SpatialDropout1D] --> O[Dense 64 + do]
%     O[Dense 64 + do] --> P[LayerNorm]
%     P[LayerNorm] --> Q[BranchDropout]
\begin{figure*}
    \centering
    \includegraphics[width=0.8\textwidth]{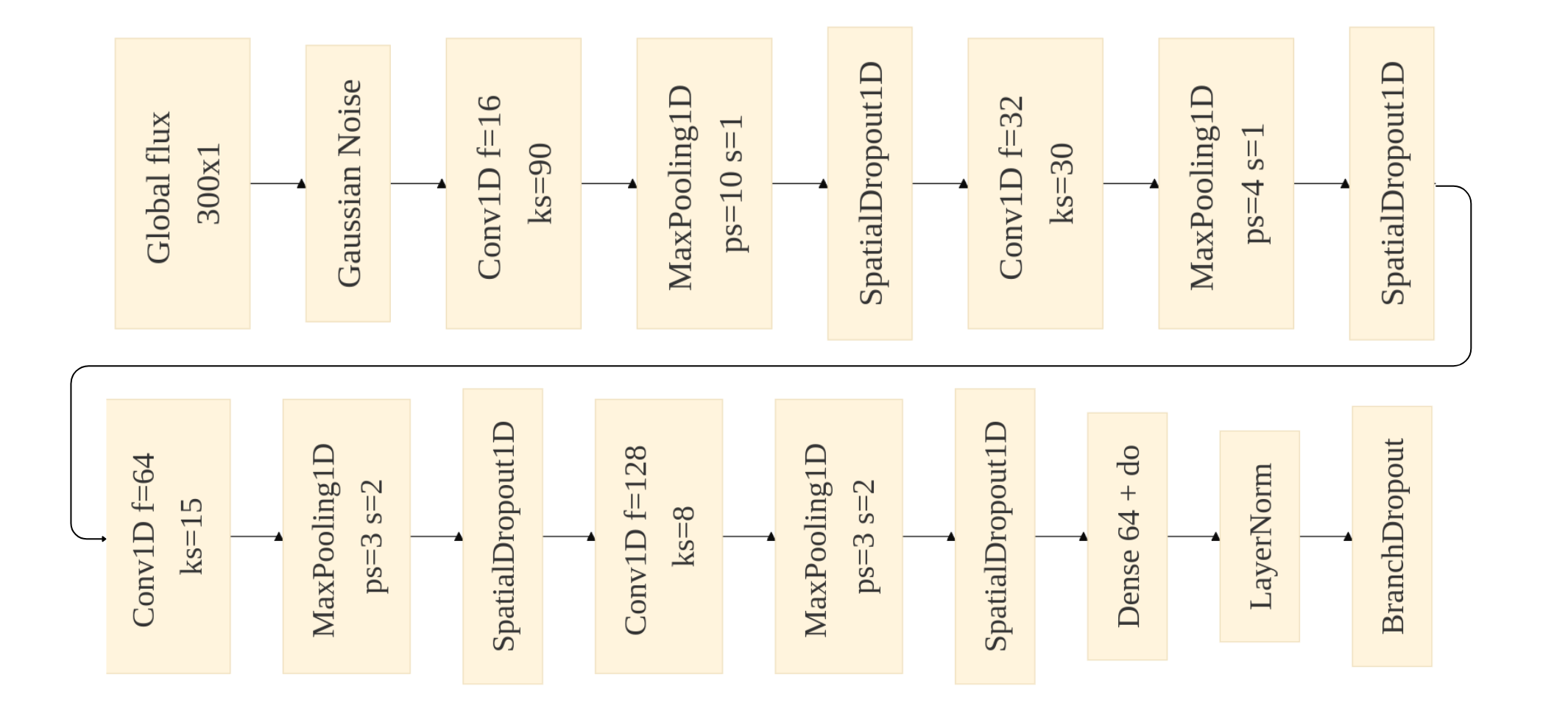}
    \caption{Neural network global flux convolutional branch. Convolutional layers have two parameters: $f$ for the number of filters and $ks$ for the kernel size. MaxPooling1D layers have two parameters $ps$ for pooling size and $s$ for strides. Dense layers include $do$ annotation for a posterior dropout layer.}
    \label{fig:global-flux-conv}
\end{figure*}

% %%{ init: {
%     'flowchart': { 'curve': 'stepBefore' },
%     'theme': 'base',
%     'themeVariables':
%         { 'fontSize': '30px', 'fontFamily': 'Inter'}
% } }%%
%flowchart TD
%    A[Numerical Params 6x1] --> G[Concatenate]
%    A1[Time Series 75xc] --> B[Gaussian Noise]
%    B[Gaussian Noise] --> C[Conv1D f=32, ks=20]
%    C[Conv1D f=32, ks=20] --> D[MaxPooling1D ps=5, s=1]
%    D[MaxPooling1D ps=5, s=1] --> E[SpatialDropout1D]
%    E[SpatialDropout1D]--> C1[Conv1D f=64, ks=10]
%    C1[Conv1D f=64, ks=10] --> D1[MaxPooling1D ps=5, s=1]
%    D1[MaxPooling1D ps=5, s=1] --> E1[SpatialDropout1D]
%    E1[SpatialDropout1D]--> C2[Conv1D f=128, ks=5]
%    C2[Conv1D f=128, ks=5] --> D2[MaxPooling1D ps=3, s=1]
%    D2[MaxPooling1D ps=3, s=1] --> F[Flatten]
%    F[Flatten] --> G[Concatenate]
%    G[Concatenate] --> H[Dense 64]
%    H[Dense 64] --> I[LayerNorm]

\begin{figure*}
    \centering
    \includegraphics[width=0.65\textwidth]{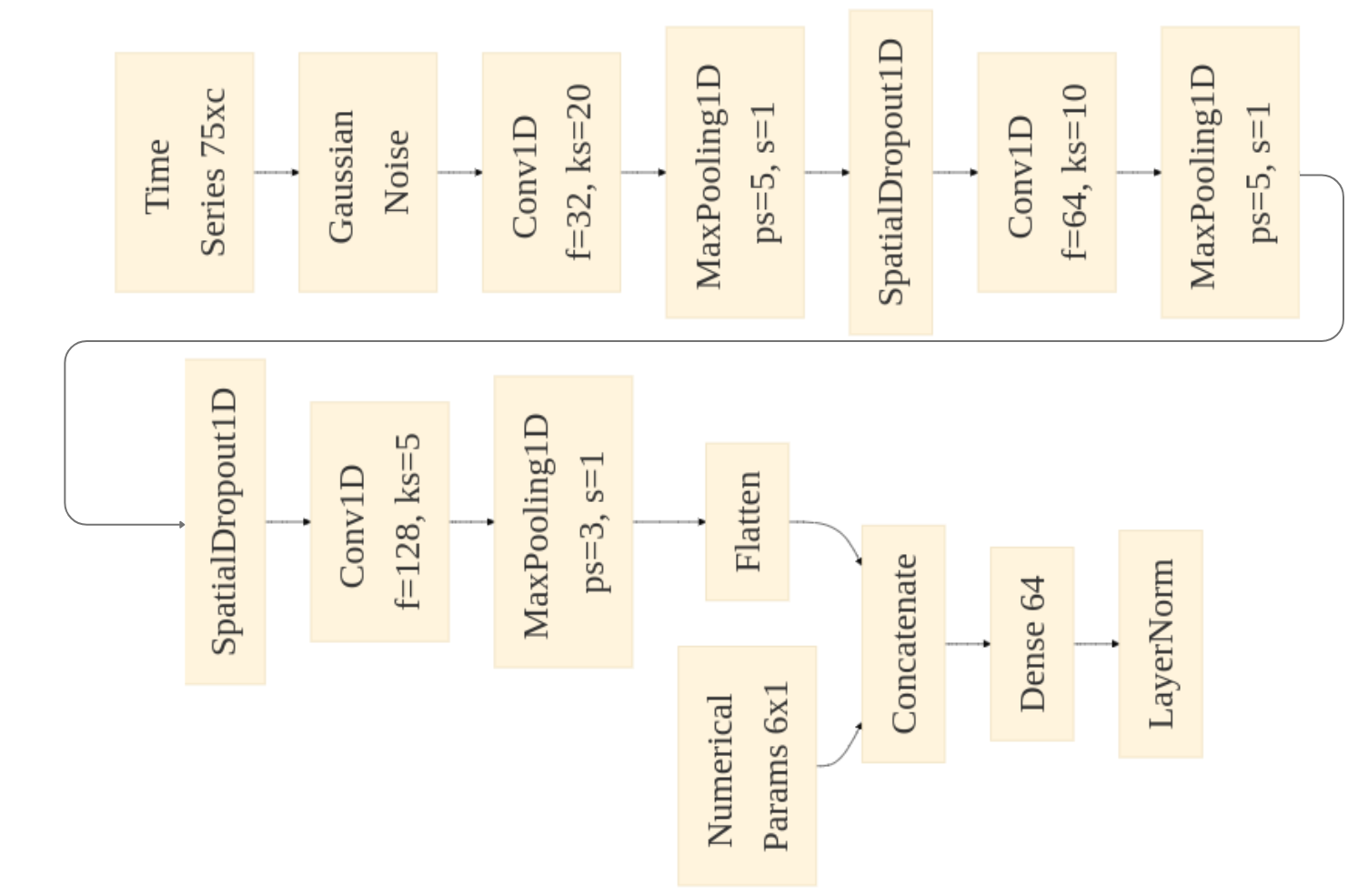}
    \caption{Neural network convolutional branch structure for focused flux data. The initial Time Series input has a size of 75x$c$, where $c$ is the number of channels of the series. The Numerical Params input (where $n$ is the number of float values it contains) is optional and is only concatenated if it is set. Convolutional layers have two parameters: $f$ for the number of filters and $ks$ for the kernel size. MaxPooling1D layers have two parameters $ps$ for pooling size and $s$ for strides. Dense layers include $do$ annotation for a posterior dropout layer.}
    \label{fig:watson-conv-branch}
\end{figure*}

% %%{ init: {
%     'flowchart': { 'curve': 'stepBefore' },
%     'theme': 'base',
%     'themeVariables':
%         { 'fontSize': '30px', 'fontFamily': 'Inter'}
% } }%%
% flowchart TD
%    L1[Dense 1000 + do] --> L2[Dense 250 + do]
%    L2[Dense 250 + do] --> L3[Dense 75 + do]
%    L3[Dense 75 + do] --> L4[Dense 35 + do]
%    L4[Dense 35 + do] --> L5[Dense 16 + do]
%    L5[Dense 16 + do] --> L6[Dense 1 + do]
%    L6[Dense 1 + do] --> L7[Sigmoid activation]

\begin{figure*}
    \centering
    \includegraphics[width=0.5\textwidth]{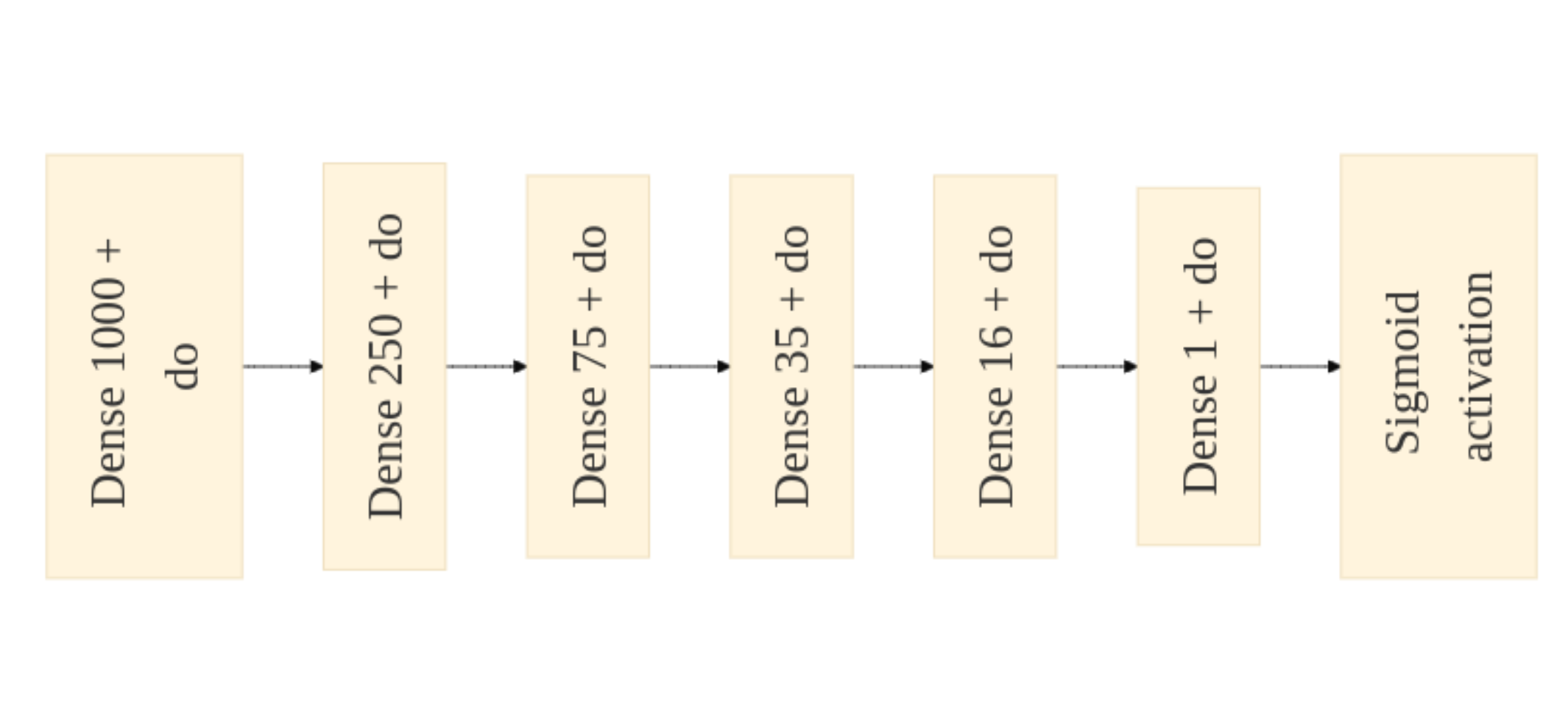}
    \caption{Neural network Feed-Forward final layers (FFN). Dense layers include $do$ annotation for a posterior dropout layer.}
    \label{fig:watson-ffn}
\end{figure*}

\begin{table}[htp]
    \centering
    \caption{Final hyperparameters selected values for the cross-validation training.}
    \label{tab:hyperparameters}
    \begin{tabular}{lr} 
        \hline  
        Hyper-parameter & Value\\
         \hline
          Epochs & 300\\
          Batch size & 200\\
          Loss function & Binary cross-entropy\\
          Class sampling & 1 \& 1\\
          Class weights & 1 \& 1\\
          Initial learning rate & $3\times10^{-4}$\\
          Epoch linear learning rate decrease & $10^{-8}$\\
          FFN Learning rate progression & 0.9\\
          Gradient clip norm. & 0.01\\
          Gradient clip value & 0.05\\
          Dropout rate & 0.2\\
          Spatial dropout rate & 0.1\\
          Branch Dropout rate & 0.05\\
          White noise std. & 0.001\\
          Dense layer L1 regularization & 0.0005\\
          Dense layer L2 regularization & 0.0005\\
          Conv. layer L1 regularization & 0.00001\\
          Conv. layer L2 regularization & 0.00001\\
          Stochastic weight avg. wait epochs & 225\\
         \hline
    \end{tabular}
\end{table}

\section{Additional metrics and predictions explanations}

\begin{figure*}
    \centering
    \includegraphics[width=0.33\textwidth]{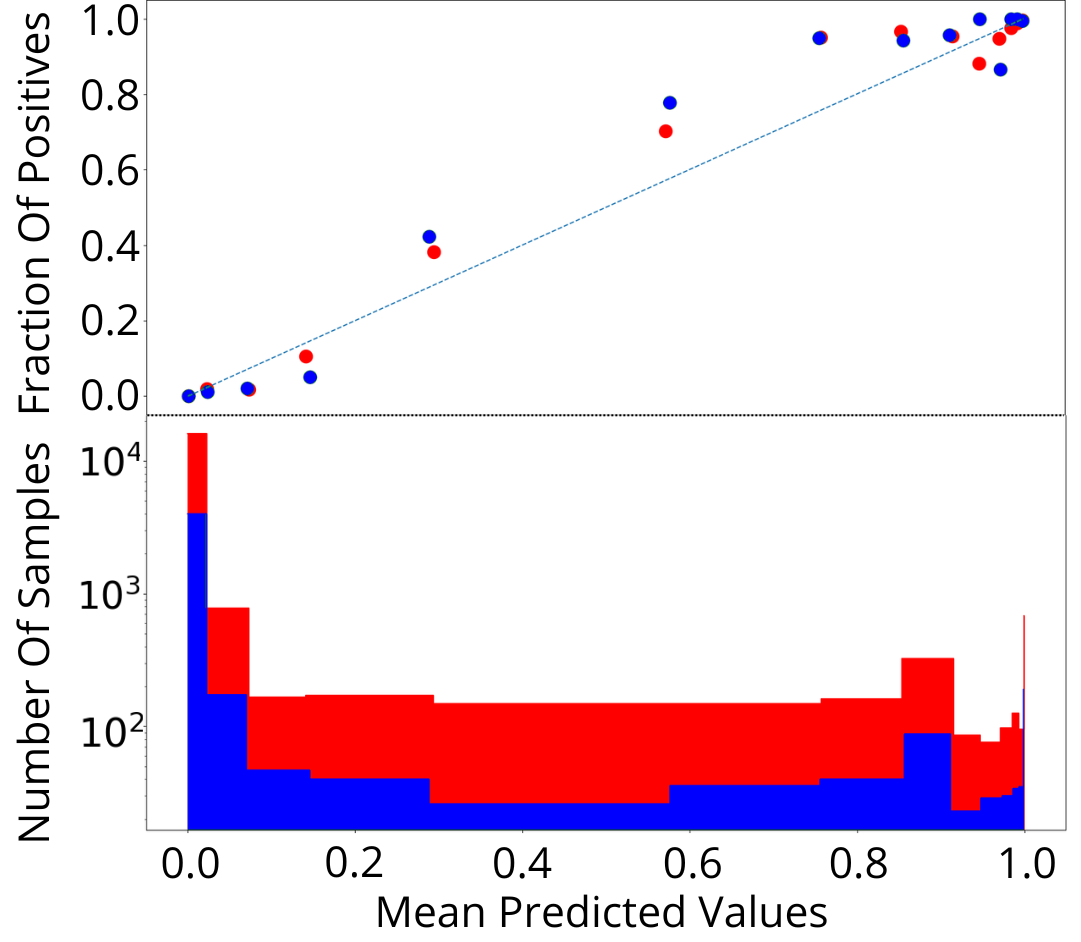}
    \includegraphics[width=0.33\textwidth]{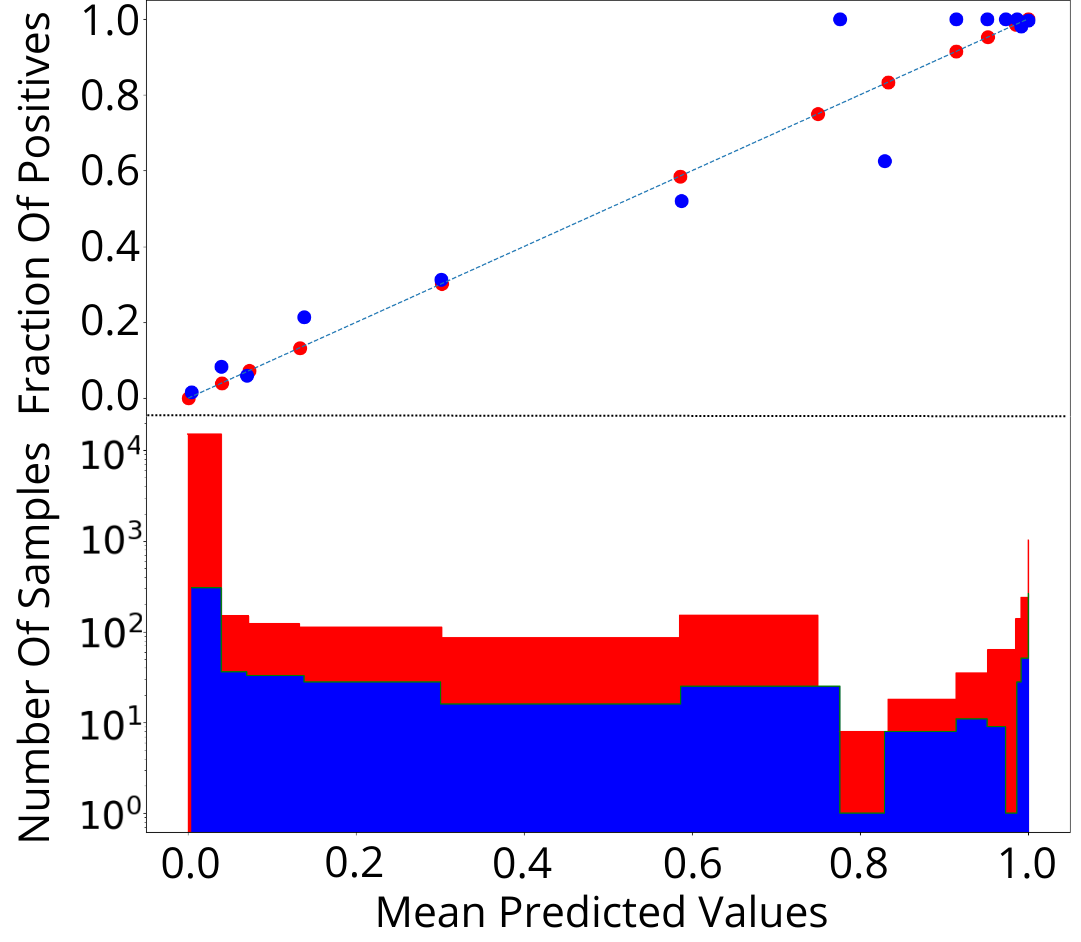}
    \includegraphics[width=0.33\textwidth]{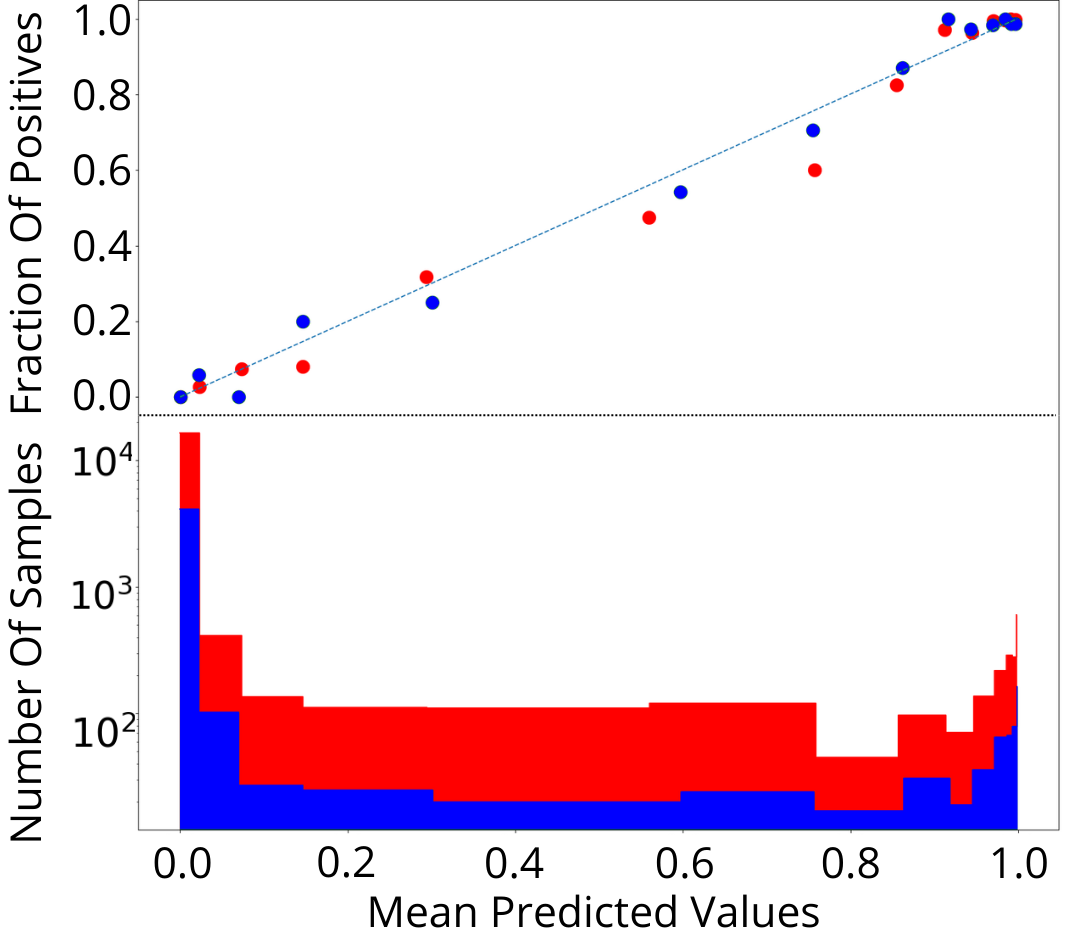}
    \caption{Calibration plot for the uncalibrated (panel left), isotonic (middle panel), and Platt (right panel) final predictions, for the entire range of predictions [0, 1]. The top panel displays the mean predicted values per bin and the bottom panel shows the number of samples per predicted values bin. The training data values are represented in red, and the validation data values are in blue. The optimal prediction likelihood line is plotted in blue.}
    \label{fig:calibrations}
\end{figure*}

\begin{table*}
    \centering
    \caption{Metrics computed for each of the trained cross-validation folds and their mean, median and standard deviation.}
    \label{tab:cv-metrics}
    \begin{tabular}{lcccccccr} 
        \hline  
        CV Fold No. & R@P0.99 & T@P0.99 & R@P1.0 & T@P1.0 & S@NPV999 & T@NPV999 & S@NPV1.0 & T@NPV1.0\\
          \hline
          1 & 0.909 & 0.869 & 0.769 & 0.954 & 0.989 & 0.160 & 0.934 & 0.009\\
          2 & 0.960 & 0.813 & 0.867 & 0.990 & 0.990 & 0.045 & 0.987 & 0.019\\
          3 & 0.897 & 0.622 & 0.332 & 0.974 & 0.988 & 0.247 & 0.000 & 0.000\\
          4 & 0.933 & 0.665 & 0.619 & 0.990 & 0.960 & 0.004 & 0.000 & 0.000\\
          5 & 0.955 & 0.463 & 0.841 & 0.729 & 0.949 & 0.024 & 0.906 & 0.012\\
          6 & 0.808 & 0.464 & 0.718 & 0.619 & 0.953 & 0.079 & 0.798 & 0.007\\
          7 & 0.861 & 0.941 & 0.368 & 0.999 & 0.977 & 0.012 & 0.000 & 0.000\\
          8 & 0.827 & 0.982 & 0.707 & 0.994 & 0.000 & 0.000 & 0.000 & 0.000\\
          9 & 0.928 & 0.978 & 0.791 & 0.999 & 0.972 & 0.003 & 0.000 & 0.000\\
          10 & 0.950 & 0.568 & 0.222 & 0.870 & 0.983 & 0.148 & 0.968 & 0.098\\
         \hline
        \multicolumn{9}{c}{Isotonic calibrated model} \\
        \hline
          1 & 0.909 & 0.715 & 0.777 & 0.942 & 0.989 & 0.272 & 0.934 & 0.014\\
          2 & 0.960 & 0.546 & 0.871 & 0.938 & 0.987 & 0.142 & 0.987 & 0.142\\
          3 & 0.897 & 0.648 & 0.341 & 0.985 & 0.987 & 0.221 & 0.780 & 0.002\\
          4 & 0.933 & 0.601 & 0.623 & 0.983 & 0.959 & 0.046 & 0.915 & 0.010\\
          5 & 0.955 & 0.236 & 0.841 & 0.962 & 0.944 & 0.037 & 0.908 & 0.012\\
          6 & 0.808 & 0.644 & 0.718 & 0.955 & 0.953 & 0.055 & 0.799 & 0.004\\
          7 & 0.853 & 0.765 & 0.390 & 0.991 & 0.973 & 0.057 & 0.000 & 0.000\\
          8 & 0.827 & 0.751 & 0.712 & 0.961 & 0.952 & 0.033 & 0.800 & 0.004\\
          9 & 0.855 & 0.667 & 0.728 & 0.938 & 0.000 & 0.000 & 0.000 & 0.000\\
          10 & 0.946 & 0.501 & 0.222 & 0.994 & 0.980 & 0.103 & 0.969 & 0.041\\
         \hline
        \multicolumn{9}{c}{Platt calibrated model} \\
        \hline
          1 & 0.909 & 0.874 & 0.773 & 0.963 & 0.989 & 0.101 & 0.931 & 0.003\\
          2 & 0.960 & 0.774 & 0.867 & 0.973 & 0.990 & 0.096 & 0.987 & 0.052\\
          3 & 0.897 & 0.756 & 0.280 & 0.998 & 0.988 & 0.213 & 0.000 & 0.000\\
          4 & 0.933 & 0.733 & 0.619 & 0.985 & 0.961 & 0.019 & 0.901 & 0.001\\
          5 & 0.955 & 0.722 & 0.837 & 0.948 & 0.946 & 0.005 & 0.879 & 0.001\\
          6 & 0.808 & 0.820 & 0.718 & 0.944 & 0.952 & 0.035 & 0.000 & 0.000\\
          7 & 0.861 & 0.881 & 0.377 & 0.992 & 0.977 & 0.048 & 0.000 & 0.000\\
          8 & 0.827 & 0.878 & 0.707 & 0.923 & 0.951 & 0.020 & 0.794 & 0.001\\
          9 & 0.928 & 0.792 & 0.800 & 0.945 & 0.972 & 0.029 & 0.000 & 0.000\\
          10 & 0.950 & 0.842 & 0.199 & 0.996 & 0.983 & 0.048 & 0.968 & 0.017\\
         \hline
    \end{tabular}
\end{table*}

\begin{table*}
    \centering
    \caption{False positives found on the combined dataset above the R@P0.99 threshold of 0.646 by the isotonic model. The horizontal line separates targets located within the VP (above) and LP (below) thresholds.}
    \label{tab:false-positives}
    \begin{tabular}{lcccl} 
        \hline  
        Target & P (d) & Original Tag & Prediction & Explanation\\
         \hline
        KIC 11760959 & 3.47 & fp & 0.993789 & Low S/N centroids shift \\
        KIC 8869680 & 7.03 & fp & 0.990741 & Computed source offset was not significative  \\
        KIC 5544450 & 4.29 & fa & 0.984733 & No worrisome parameters found for rejection \\
        KIC 2446113 & 6.72 & fp & 0.984733 & Computed source offset was not significative \\
        KIC 7732285 & 4.39 & fp & 0.982456 & Low S/N centroids shift, no source offset spotted \\
        KIC 7594098 & 171.43 & tce & 0.961538 & No worrisome parameters for rejection \\
        \hline  
        KIC 8681125 & 153.78 & tce\_odd\_even & 0.960000 & Even transits are plain, no explanation for rejection \\
        KIC 3542117 & 12.58 & fp & 0.954545 & Source offset slightly significant, but insufficient for rejection \\
        KIC 6933567 & 4.54 & tce\_source\_offset & 0.941176 & Computed source offset was not significative enough \\
        KIC 5297298 & 34.19 & fp & 0.941176 & 22 $R_\oplus$ radius did not trigger rejection, low S/N centroids shift \\
        KIC 11392618 & 110.92 & fp & 0.937500 & Computed source offset was not significative  \\
        KIC 7767559 & 4.41 & fp & 0.937500 & Noisy signal with harmonic TCE too, no visual data for rejection \\
        KIC 6290935 & 3.42 & tce & 0.937500 & No worrisome parameters found for rejection \\
        KIC 7375795 & 19.30 & fp & 0.928571 & Computed source offset was not significative  \\
        KIC 8564976 & 152.83 & fp & 0.875000 & Significant but noisy centroids shift \\
        KIC 5632093 & 0.50 & tce & 0.764706 & Noisy signal with harmonic TCE too, no visual data for rejection \\
        KIC 5524881 & 7.20 & fp & 0.764706 & Computed source offset was not significative enough  \\
        KIC 2573108 & 5.09 & fa & 0.764706 & Low S/N signal and odd/even effect were not enough for rejection\\
        KIC 3533469 & 10.22 & fp & 0.764706 & Low S/N source offset and centroids shift  \\
        KIC 2989706 & 1.26 & fp & 0.750000 & Low S/N centroids shift \\
        KIC 8804455 & 2.39 & fp & 0.750000 & Optical ghost and centroids shift were significant but noisy\\
        KIC 5283542 & 5.98 & fp & 0.750000 & Low S/N source offset and centroids shift\\
        KIC 7222086 & 2.44 & fp & 0.714286 & The V-shape of the signal is not enough to trigger a rejection \\
        KIC 8261920 & 15.45 & tce & 0.714286 & No worrisome parameters found for rejection\\
        KIC 8655354 & 3.36 & tce & 0.692308 & Centroids shift not significative enough\\
        KIC 6889235 & 2.59 & tce\_centroids\_offset & 0.692308 & Centroids shift and optical ghost were significant but noisy\\
        KIC 9873759 & 4.99 & fp & 0.692308 & No worrisome parameters found for rejection\\
        KIC 4557777 & 1.87 & tce\_og & 0.692308 & Extremely noisy and variable light curve\\
        KIC 12021943 & 6.10 & fp & 0.666667 & No worrisome parameters for rejection\\
        KIC 10399321 & 21.58 & fp & 0.647059 & Low S/N source offset, centroids shift and optical ghost \\
        KIC 4936524 & 1.67 & tce & 0.647059 & Noisy signal with no worrisome parameters for rejection\\
        KIC 6699368 & 3.69 & fp & 0.647059 & Computed source offset not significative  \\
        KIC 9770983 & 7.03 & fp & 0.647059 & No worrisome parameters found for rejection\\
        KIC 7767559 & 2.20 & tce & 0.647059 & Noisy signal with harmonic TCE too, no visual data for rejection \\
        KIC 8260269 & 6.71 & fp & 0.647059 & Computed source offset not significative enough\\
        \hline
    \end{tabular}
\end{table*}

\begin{table*}
    \centering
    \caption{False negatives found on the combined dataset above the S@NPV999 threshold of 0.056 by the isotonic models. The horizontal line separates targets located within the VN (above) and LN (below) thresholds. The DR-25 metric column represents the SPOC DVR values that were marked as problematic.}
    \label{tab:false-negatives}
    \begin{tabular}{lcccl}
        \hline  
        Target & P (d) & DR-25 metric & Prediction & Explanation\\
        \hline
        KIC 11446443 & 2.47 & Centroids shift & 0.002232 & Slight secondary eclipse, strong centroids shift\\
        KIC 10989274 & 16.18 & Source offset & 0.004202 & Ultra-noisy light curve\\
        KIC 1718189 & 13.06 & Centroids significance & 0.004630 & Slight source offset and centroids shift\\
        \hline
        KIC 10937029 & 328.24 & Source offset, centroids shift & 0.004762 & Optical ghost and centroids shift data are significative\\
        KIC 8873450 & 24.28 & Optical ghost & 0.006061 & Slightly noisy baseline, no clear reason for rejection\\
        KIC 11521793 & 16.01 & - & 0.008403 & Slight secondary eclipse\\
        KIC 8311864 & 384.85 & Source offset & 0.010417 & Computed source offset was significative\\
        KIC 9715631 & 425.48 & Model Chi Square & 0.010638 & Significant source offset\\
        KIC 11027624 & 394.62 & - & 0.012987 & In-transit data was missing.\\
        KIC 10397751 & 3.45 & - & 0.014925 & Low computed transit S/N\\
        KIC 7703955 & 265.48 & Centroids significance & 0.018519 & Slight secondary event, no clear reason for rejection\\
        KIC 8410727 & 25.27 & - & 0.029412 & High odd/even stat, noisy centroids shift\\
        KIC 10187017 & 5.29 & Source offset & 0.033333 & Noisy data, odd/even, probably wrong ephemeris\\
        KIC 9141746 & 2.85 & - & 0.037736 & Low S/N signal\\
        KIC 8056665 & 84.69 & Centroids shift & 0.037736 & Improperly normalized global light curve\\
        KIC 8395660 & 43.84 & - & 0.041667 & Improperly normalized global light curve\\
        KIC 3645438 & 386.37 & - & 0.041667 & Slight optical ghost and centroids shift significance \\
        KIC 11017901 & 7.79 & Source offset, centroids shift & 0.046512 & Computed source offset was significative\\
        KIC 8891684 & 374.88 & Optical ghost & 0.047619 & Improperly normalized secondary transit view\\
        KIC 10397751 & 2.01 & - & 0.047619 & Low S/N transit, slight source offset\\
        KIC 10028792 & 191.23 & - & 0.047619 & Slight optical ghost and centroids shift\\
        KIC 7100673 & 2.89 & - & 0.055556 & Improperly normalized secondary transit view\\
        KIC 4735826 & 7.55 & Target magnitude & 0.055556 & Slight odd/even difference, low computed transit S/N\\
        KIC 6929841 & 3.52 & - & 0.055556 & No worrisome data for acceptance\\
        KIC 9141355 & 469.62 & Centroids shift & 0.055556 & Slightly significant source offset\\
        KIC 11138155 & 4.96 & Centroids shift & 0.055556 & Slight significance of centroids shift\\
        \hline
    \end{tabular}
\end{table*}

\begin{table*}
    \centering
    \caption{False positives found on the TESS test set. The horizontal line separates targets located within the VP (above) and LP (below) thresholds.}
    \label{tab:tess-false-positives}
    \begin{tabular}{lccl} 
        \hline  
        Target & P (d) & Prediction & Explanation\\
         \hline
        TIC 300013921 & 54.66 & 0.996926 & Unclear acceptance. Centroids shift is significant \\
        TIC 97158538 & 2.60 & 0.995455 & No problematic metrics. Star radius of 2.24~$R_\odot$ \\
        TIC 301051430 & 3.27 & 0.995172 & Unclear acceptance. Centroids shift is significant \\
        TIC 82707763 & 6.66 & 0.990626 & No problematic metrics. Slight centroids shift \\
        TIC 470974162 & 3.26 & 0.963706 & No problematic metrics. Slight centroids shift \\
        TIC 230017325 & 9.69 & 0.963665 & Unclear acceptance. Centroids shift is significant \\
        \hline
        TIC 233684822 & 4.68 & 0.954010 & Unclear acceptance. Centroids shift and optical ghost are significant \\
        TIC 55383975 & 48.05 & 0.936948 & Unclear acceptance. Centroids shift and optical ghost are significant \\
        TIC 99330090 & 7.01 & 0.926981 & No problematic metrics \\
        TIC 54044474 & 3.70 & 0.919622 & Unclear acceptance. Odd/even, optical ghosts are significant, star radius of 1.90~$R_\odot$ \\
        TIC 220016044 & 0.94 & 0.917404 & No problematic metrics. High bootstrap FAP, slight centroids shift. \\
        TIC 289988797 & 2.20 & 0.915701 & No problematic metrics \\
        TIC 363761216 & 7.41 & 0.914659 & Unclear acceptance. Centroids shift and optical ghost are significant \\
        TIC 272050287 & 8.50 & 0.902482 & No problematic metrics \\
        TIC 229976631 & 10.72 & 0.890349 & No problematic metrics \\
        TIC 376981340 & 4.02 & 0.885384 & No problematic metrics. Slight centroids shift \\
        TIC 170103435 & 1.98 & 0.864457 & Unclear acceptance. Centroids shift is significant \\
        TIC 470709084 & 3.67 & 0.842780 & Unclear acceptance. Centroids shift is significant \\
        TIC 90448944 & 1.44 & 0.826802 & Unclear acceptance. Centroids shift is significant \\
        TIC 379723086 & 7.44 & 0.801629 & Unclear acceptance. Centroids shift is significant \\
        TIC 271581073 & 2.67 & 0.783800 & Unclear acceptance. Source offset is significant \\
        TIC 219208037 & 0.98 & 0.775675 & No problematic metrics \\
        TIC 69672493 & 1.49 & 0.775388 & No problematic metrics. Slight centroids shift \\
        TIC 302659412 & 14.52 & 0.750650 & Unclear acceptance. Centroids shift is significant \\
        TIC 382437043 & 14.64 & 0.729115 & Unclear acceptance. Centroids shift and source offset are significant \\
        TIC 316937670 & 0.62 & 0.718872 & Unclear acceptance. Source offset is significant \\
        TIC 169904935 & 4.55 & 0.715511 & Unclear acceptance. Centroids shift is significant \\
        TIC 290302097 & 2.99 & 0.706066 & Unclear acceptance. Centroids shift and source offset are significant \\
        TIC 151959065 & 10.54 & 0.705646 & No problematic metrics. Slight centroids shift \\
        TIC 231721005 & 9.02 & 0.695996 & Unclear acceptance. Centroids shift is significant \\
        TIC 289535142 & 2.34 & 0.692081 & No problematic metrics. Slight odd/even effect \\
        TIC 169532369 & 3.60 & 0.671492 & No problematic metrics. Slight source offset \\
        \hline
    \end{tabular}
\end{table*}

\begin{table*}
    \centering
    \caption{False negatives found on the TESS test set. The horizontal line separates targets located within the LN (above) and VN (below) thresholds.}
    \renewcommand{\arraystretch}{0.88} % Reduce row height
    \label{tab:tess-false-negatives}
    \begin{tabular}{lccl} 
        \hline  
        Target & P (d) & Prediction & Explanation\\
         \hline
        TIC 267574918 & 1.41 & 0.000010 & Wrong transit duration, centroids shift detected \\
        TIC 380619414 & 2.66 & 0.000010 & Wrong stellar radius, centroids shift detected \\
        TIC 138359318 & 1.02 & 0.000010 & Wrong stellar radius, centroids shift detected \\
        TIC 158324245 & 1.76 & 0.000010 & Wrong stellar radius, secondary transit and centroids shift detected \\
        TIC 164786087 & 4.08 & 0.000010 & Wrong stellar radius, slight centroids shift \\
        TIC 166527623 & 6.96 & 0.000010 & Noisy data \\
        TIC 329148988 & 27.27 & 0.000010 & Low S/N signal \\
        TIC 404243877 & 3.20 & 0.000010 & Noisy data, slight centroids shift \\
        TIC 164173105 & 3.07 & 0.000010 & Noisy data \\
        TIC 363548415 & 6.77 & 0.000643 & Low S/N signal \\
        TIC 142276270 & 26.32 & 0.001064 & Secondary transit and slighly imprecise ephemeris \\
        TIC 175236511 & 7.81 & 0.001287 & Low S/N signal\\
        TIC 35516889 & 0.79 & 0.001712 & Secondary transit and centroids shift detected \\
        TIC 27491137 & 10.36 & 0.002805 & Slight secondary transit, source offset detected \\
        TIC 392476080 & 0.67 & 0.004059 & Secondary transit,  planet radius R~=~16$R_\oplus$ \\
        TIC 16740101 & 1.48 & 0.004251 & Secondary transit, planet radius R~=~20$R_\oplus$ \\
        TIC 34068865 & 0.32 & 0.004704 & Wrong transit duration \\
        TIC 37718056 & 3.66 & 0.005023 & Imprecise ephemeris \\  
        TIC 424865156 & 2.20 & 0.005886 & Secondary transit, planet radius R~=~17$R_\oplus$, centroids shift \\
        TIC 134537478 & 1.71 & 0.006542 & Secondary transit, V-shape, centroids shift, optical ghost \\
        TIC 129979528 & 1.22 & 0.006665 & Secondary transit, centroids shift, planet radius R~=~18$R_\oplus$ \\
        \hline
        TIC 22529346 & 1.27 & 0.007925 & Secondary transit, centroids shift, source offset, planet radius R~=~20$R_\oplus$ \\  
        TIC 259506033 & 1.9 & 0.009917 & Imprecise ephemeris \\ 
        TIC 116264089 & 1.31 & 0.013275 & Centroids shift, source offset, planet radius R~=~17$R_\oplus$ \\
        TIC 257527578 & 54.32 & 0.014233 & Missing in-transit data \\ 
        TIC 339522221 & 1.82 & 0.016944 & Imprecise ephemeris \\
        TIC 55652896 & 34.51 & 0.017319 & Wrong transit duration \\  
        TIC 267572272 & 2.80 & 0.018388 & Noisy data \\  
        TIC 7088246 & 2.16 & 0.020782 & Imprecise ephemeris, planet radius R~=~23$R_\oplus$ \\   
        TIC 48217457 & 3.12 & 0.020237 & Centroids shift, planet radius R~=~16$R_\oplus$\\  
        TIC 142276270 & 18.80 & 0.021948 & Noisy data, slightly imprecise ephemeris \\  
        TIC 427761355 & 1.90 & 0.023950 & Secondary transit, planet radius R~=~20$R_\oplus$\\  
        TIC 232568235 & 0.35 & 0.024026 & Ultra-short signal \\  
        TIC 236445129 & 0.97 & 0.025096 & Secondary transit, centroids shift, planet radius R~=~17$R_\oplus$ \\  
        TIC 14614418 & 0.84 & 0.025297 & Centroids shift, planet radius R~=~16$R_\oplus$ \\  
        TIC 86396382 & 1.09 & 0.025896 & Secondary transit, centroids shift, planet radius R~=~24$R_\oplus$ \\ 
        TIC 466840711 & 2.49 & 0.029619 & Planet radius R~=~36$R_\oplus$ \\
        TIC 436575927 & 5.1 & 0.030465 & High bootstrap FAP \\  
        TIC 365007485 & 33.59 & 0.032347 & Odd view missing \\
        TIC 4610830 & 8.26 & 0.033315 & Low S/N, missing odd, slight centroids shift, low quality transits, high bootstrap FAP \\  
        TIC 432549364 & 1.22 & 0.034280 & Blended light curve, secondary transit \\  
        TIC 240681314 & 2.73 & 0.035407 & V-shaped transit, centroids shift, planet radius R~=~18$R_\oplus$ \\  
        TIC 330690135 & 14.76 & 0.035805 & Low good quality transits, high bootstrap FAP, slightly imprecise ephemeris \\  
        TIC 293435336 & 1.81 & 0.035325 & Secondary transit, centroids shift, planet radius R~=~22$R_\oplus$ \\  
        TIC 354619337 & 2.15 & 0.036172 & Secondary transit, planet radius R~=~17$R_\oplus$ \\  
        TIC 17865622 & 3.12 & 0.036560 & Source offset\\ 
        TIC 178284730 & 2.24 & 0.040843 & Centroids shift, source offset \\
        TIC 251848941 & 20.71 & 0.041255 & Low S/N signal, low good quality transits \\  
        TIC 366411016 & 8.87 & 0.041871 & Slightly high FAP, slight centroids shift\\
        TIC 150096001 & 11.02 & 0.042953 & Low good quality transits \\         
        TIC 137402055 & 1.97 & 0.043426 & Slight odd/even effect, centroids shift \\
        TIC 388811663 & 9.8 & 0.044150 & Low S/N signal, high FAP\\  
        TIC 55092869 & 4.74 & 0.044605 & Noisy centroids and optical ghost views, high bootstrap FAP \\    
        TIC 245392284 & 1.74 & 0.050155 & Centroids shift, source offset \\
        TIC 176868951 & 0.37 & 0.053346 & Slight centroids shift \\     
        \hline
    \end{tabular}
\end{table*}

\end{appendix}

\end{document}